\documentclass[aps,pre,notitlepage,nofootinbib,superscriptaddress]{revtex4}
\usepackage{graphicx}
\usepackage{epstopdf}
\usepackage{bm}
\usepackage{amssymb}
\usepackage{psfrag}
\usepackage{color}
\usepackage{amsbsy}      
\usepackage{amsmath,amsthm}
\usepackage{mathtools}
\usepackage{dsfont}
\usepackage{makecell}
\usepackage{slashbox}
\usepackage{lineno}
\usepackage{comment}
\usepackage{hyperref}
\usepackage{tcolorbox}
\usepackage{soul,color}
\usepackage{url}
\usepackage{etoolbox}
\patchcmd{\section}
  {\centering}
  {\raggedright}
  {}
  {}
\patchcmd{\subsection}
  {\centering}
  {\raggedright}
  {}
  {}
\graphicspath{{./figs/}}
\usepackage{newfloat}
\DeclareFloatingEnvironment[name={Fig. D}]{suppfigure}

\usepackage{natbib} 

\usepackage{pifont}% http://ctan.org/pkg/pifont

\def\a{{\bf a}}
\def\b{{\bf b}}
\def\W{{\bf W}}

\def\dd{\textrm{d}}
\def\C{{\bf C}}

\def\x{{\bf x}}
\def\y{{\bf y}}

\def\f{\frac}

\def\p{\partial}

\def\A{{\bf A}}
\def\B{{\bf B}}
\def\C{{\bf C}}

\def\L{{\bf \Lambda}}

\def\W{{\bf W}}
\def\X{{\bf X}}

\def\Z{{\bf Z}}
\def\Y{{\bf Y}}

\def\p{\partial}

\definecolor{red}{rgb}{0.7,0,0}
\definecolor{green}{rgb}{0,0.55,0}

\usepackage{dsfont}
\usepackage{enumerate}
\begin{document}

\title{Kinetic theory for
  structured populations: application to stochastic sizer-timer models
  of cell proliferation}

\author{Mingtao Xia$^{1}$ and Tom Chou}

\affiliation{Department of Mathematics, UCLA, Los Angeles, CA,
  90095-1555, USA}
\affiliation{Department of Computational Medicine, UCLA, Los Angeles, CA,
  90095-1766, USA}

\email{tomchou@ucla.edu}


%\maketitle

%\runninglinenumbers

\begin{abstract}
  We derive the full kinetic equations describing the evolution of the
  probability density distribution for a structured population such as
  cells distributed according to their ages and sizes.  The kinetic
  equations for such a ``sizer-timer" model incorporates both
  demographic and individual cell growth rate stochasticities.
  Averages taken over the densities obeying the kinetic equations can
  be used to generate a second order PDE that incorporates the growth
  rate stochasticity. On the other hand, marginalizing over the
  densities yields a modified birth-death process that shows how age
  and size influence demographic stochasticity. Our kinetic framework
  is thus a more complete model that subsumes both the deterministic
  PDE and birth-death master equation representations for structured
  populations.
\end{abstract}

%
%Uncomment for keywords
%\noindent{\it Keywords}: Age Structure, Birth-Death Process, Kinetics,
%Fission

%\submitto{Physical Biology}
%\maketitle


\maketitle
\noindent{\it Keywords}: Age Structure, Birth-Death Process, Kinetics,
Fission

\section{Introduction}

Across many diverse applications, mathematical models have been
formulated to describe the evolution of populations according to a
number of individual attributes such as age, size, and/or added size
since birth.  For example, deterministic age-structured models that
incorporate age-dependent birth and death were developed by McKendrick
and have been applied to human populations \cite{von1959some}. More
recently, there has been renewed interest in cell size control
\cite{JUN2015,burov2018effective}, cellular division mechanisms
\cite{DOUMIC2014}, and structured cell population models
\cite{PERTHAME2008,Metz1986}.

When considering proliferating cell populations, individual cell
growth is interrupted by cell division events that generate smaller
daughter cells. Cell division is a process that involves many
biochemical steps and complex biophysical mechanisms that involves
metabolism, gene expression, protein production, DNA replication,
chromosome separation (for eukaryotic cells), and fission or cell wall
formation
\cite{MAALOE1973,huisman1981inducible,CHANDLER_BROWN,OSKAR2017,
  wessels1994developmental}.  To simplify the understanding of which
factors trigger cell division, three basic models that subsume these
complex processes have been proposed.  Cells can divide based on their
age since birth, volume (size), or added volume since birth $y$
\cite{JUN2015,SINGH2017}.  PDE approaches for the timer,
sizer, and adder models, as well as combinations of these models, have
been well-studied \cite{Xia2020,Metz1986,BERNARD2016}. These PDE
approaches implicitly describe the mean density of cells in age, size,
and/or added size, and are considered deterministic models.
%Their derivation assumes deterministic growth dynamics of individual
%cells.  For example, the size $x(t)$ between cell birth/fission might
%grow exponentially according to $\dot{x} = \lambda x$.  Such models
%have been shown to exhibit interesting behavior including blow-up of
%population-averaged cell sizes \cite{burov2018effective}.

However, there has been much less development of structured
populations models that incorporate stochastic effects.  In the
presence of stochasticity, how would the PDEs be modified? In the
sizer-timer type of structured population models, stochasticity can
arise in the growth dynamics of each cell as well as in the random
times of cell division and death (demographic stochasticity).

Stochasticity arising from random times of birth and death
(demographic stochasticity) has been considered in timer-like models
for age-structured populations
\cite{chou2016hierarchical_PRE,chou2016hierarchical}.  This approach
generalized the classic deterministic McKendrick equation to a higher
dimension (dynamically varying) associated with the number of
individuals in the system.  This higher-dimensional stochastic
``kinetic theory'' allows one to systematically connect an
age-indepedent birth-death master equation description to the
deterministic age-structured McKendrick model.  A comprehensive and
general treatment of the age-structured stochastic process using a
Doi-Peliti operator formulism has also been developed for calculation
of correlation functions \cite{Greenman_Path}. The full kinetic theory
has only been developed for age-structured populations and only
includes demographic stochasticity (since chronological age is a
deterministic quantity proportional to time).  Other approaches using
stochastic hybrid systems \cite{SINGH} have been used to incorporate
the influence of random birth times of population-level variations in
cell size.  Intrinsic stochasticity in the growth rate of an
individual cell has been treated in terms of Langevin equations for
cell size \cite{AMIR_REVIEW}, effective potentials
\cite{burov2018effective} and stochastic maps
\cite{SINGH2017,BUROV_MAP}.  Recently, Chapman-Kolmogorov equations
have also been applied to study the effect of different sources of
noise in cellular proliferation \cite{nieto2020continuous}.  However,
stochasticity in the intrinsic growth rate has not been considered
within demographically stochastic kinetic theory.

%
%Such noise in cell sizes can also arise from
%demographic stochasticity (variations in inter-birth times).
%

%This
%stochastic mechanism was found to admit ``blow-up'' in which the
%expected cell sizes can increase without bound with increasing
%generation.  Modi \textit{et al.} \cite{SINGH2017} used a different
%additive noise model for cell size changes with each generation and do
%not find blow-up even though it is known to occur in a deterministic
%PDE model of cell size distributions. These stochastic map approaches
%are not equipped to describe population-level distributions in size or
%age.

%additive noise model for cell size changes with each generation and do
%not find blow-up even though it is known to occur in a deterministic
%PDE model of cell size distributions. These stochastic map approaches
%are not equipped to describe population-level distributions in size or
%age.

%This
%stochastic mechanism was found to admit ``blow-up'' in which the
%expected cell sizes can increase without bound with increasing
%generation.  Modi \textit{et al.} \cite{SINGH2017} used a different
%additive noise model for cell size changes with each generation and do
%not find blow-up even though it is known to occur in a deterministic
%PDE model of cell size distributions. These stochastic map approaches
%are not equipped to describe population-level distributions in size or
%age.

In this paper, we shall derive a kinetic theory for the sizer-timer
model of cell proliferation that incorporates both demographic
stochasticity and intrinsic stochasticity in the growth of individual
cells. In the next section, we derive the Fokker-Planck equation for
the size of an individual cell and define the probabilistic quantities
needed to construct the full kinetic theory.  This equation is then
marginalized in Section \ref{meanfield} to explicitly isolate and show
the feature limits of intrinsic stochasticity and demographic
stochasticity. Including both sources of stochasticity renders the
calculations of marginalized densities rather technical, but by
successively taking the marginalized single-density limits, we show
how the theory reduces to simpler forms and reveal the procedure for
solving the full high-dimensional problem. Moreover, by taking higher
moments of the density, an unclosed hierarchy of equations that
reflect demographic stochasticity arises. Our results generalize a
large body of work on sizer-timer PDE models to include stochastic
processes, both at the individual and population levels.

\section{Derivation of kinetic theory}
\label{DERIVATION}

Here we outline the derivation of the kinetic equation for a
population of dividing cells of different ages $a$ and sizes (volumes)
$x$. We start from the SDE for the size~\footnote{Alternatively,
  $X_{t}$ might also represent the log of the cell size} of a
single cell at time $t$:

\begin{equation}
\dd X_t = g(X_t, A_t, t)\dd t +
\sigma(X_t, A_t, t)\dd{W_t}, \quad X_t,\, A_t\in\L,
\label{SDE0}
\end{equation}
where $\L\coloneqq[0, \infty)$, $A_t$ is the cell's age (time that has
  elapsed after its birth), $g(X_t,A_t,t)$ is the size- and
    age-dependent growth rate, and $W_t$ is a standard Wiener process
  with independent, normally distributed increments $W_t-W_s$, zero
  mean, and variance $t-s$. The parameter $\sigma(X_t, A_{t}, t)$
  represents the strength of stochasticity in cell's growth rate.
  Here, we assume both $g$ and $\sigma$ are Lipschitz continuous to
  ensure the existence and uniqueness of $X_t$ given any initial
  conditions $X_0>0, A_0 \geq 0$. We also assume $\sigma\in\C^1,
  \sigma(0, t, a)=\partial_{x}\sigma(0, t, a)=0$ so that the noise
  vanishes at $x=0$ and $X_t$ remains positive.

%The integral form to Eq.~\eqref{SDE0} is
%
%\begin{equation}
%X_t = X_{t'}+\int_{t'}^tg(X_s, A_s, s)\dd{s}+\int_{t'}^{t}\sigma(X_s,
%A_s, s)\dd{W_s},\quad t'<t.
%\label{SOLN0}
%\end{equation}
%
Next, we investigate a system of $m+2n$ cells, where $m$ is the number
of individual cells (singlets) and $n$ is the number of twins
(doublets). A twin means two daughter cells generated from the
division of a common mother cell, and therefore they have the
identical age.  In this section, we use the notation

\begin{equation}
\begin{aligned}
\X^{(m)}_t & = (X^1_t, X^2_t,..., X^m_t),\, \Y^{(2n)}_t  = (Y^1_t,..., Y^{2n}_t), \\
\A^{(m)}_t  &= (A^1_t, A^2_t,..., A^m_t),\,\, \B^{(n)}_t = (B^1_t, ..., B^n_t),
\label{RandV}
\end{aligned}
\end{equation}
where $\A^{(m)}_t$ and $\B^{(n)}_t$ are ordered ages such that
$A_t^i\geq A_t^j\geq0, B_t^i\geq B_t^j\geq0, \forall i>j$ and
$\X^{(m)}_t$ and $\Y^{(2n)}_t$ are the vectors of the volumes of the
$m$ singlets and $2n$ doublets that are of ages $\A^{(m)}_t$ and
$\B^{(n)}_t$, respectively, at time $t$.  Note that two cells in a
doublet have the same age but can have different sizes; thus, the age
vector $\B_t^{(n)}$ of the $2n$ twins stores $n$ ages, while the size
vector $\Y^{(2n)}_t$ stores $2n$ sizes.

Formally solving Eq.~\eqref{SDE0}, each $X_t^i$ and $Y_t^j$ satisfies

\begin{equation}
\begin{aligned}
X_t^i = & X_{t'}^i+\int_{t'}^tg(X_s^i, A_s^i, s)\dd{s}
+\int_{t'}^{t}\sigma(X_s,A_s, s)\dd{W_s^{i}},  \\
Y_t^j  = & Y_{t'}^j+\int_{t'}^tg(Y_s^j, B_s^{[\frac{j+1}{2}]}, s)\dd{s}
+\int_{t'}^{t}\sigma(Y_s^j, B_s^{[\frac{j+1}{2}]}, s)\dd W^{m+j}_{s},
\end{aligned}
\end{equation}
where $\dd W_s^{i}, \dd W_s^{m+j}$ are the intrinsic, indepedent
fluctuations in growth rates.  We assume that cell division rates are
regulated by a ``timer" mechanism and does not depend on cell size,
\textit{i.e.}, the probability that a cell in a population of $m$
singlets and $n$ doublets divides during $(t, t+\Delta{t}]$ is
$\beta_{m, n}(A_t,t)\dd{t} + o(\dd{t})$, a function of its age $A_t$,
time $t$ and population sizes $m, n$. The mathematical analyses that
follow require that the birth rate is independent of a cell's size
$X_t$.  Finally, we take the continuous time limit and assume that in
a finite number of cells, the possibility of two cells dividing in
$(t, t+\dd{t}]$ is $o(\dd{t})$ as $\dd{t}\rightarrow0$.

\subsection{The forward equation}
\label{forwardeqn}
We evaluate the increment in time by Ito's formula applied to a
function $f_{m, n}(\X^{(m)}_t\!\!\!,\, \Y^{(2n)}_t\!\!\!,\, t;
\A^{(m)}_{t'}\!\!\!, \,\B^{(n)}_{t'})$ of $m$ individual and $n$ twin
sizes given initial sizes and ages $\A^{(m)}_{t'}\!\!\!,
\B^{(n)}_{t'}$ at $t'<t$, where the ages are defined to be in the
descending order $A^1\geq A^2...\geq A^m \geq 0$, $B^1\geq B^2...\geq
B^n\geq0$.  Ordering the ages allows us to easily incorporate cell
division as a boundary condition in which newborn cells are represented by
$B^n=0$.

\begin{equation}
\begin{aligned}
& f_{m, n}(\X^{(m)}_{t+\dd{t}},\Y^{(2n)}_{t+\dd{t}},t+\dd{t};\A^{(m)}_{t'}\!\!\!,\,
\B^{(n)}_{t'}) -
  f_{m, n}(\X^{(m)}_{t}\!\!\!,\,\Y^{(2n)}_{t}\!\!\!,\, t; \A^{(m)}_{t'}\!\!\!\!,\, \B^{(n)}_{t'})  \\
%= \hspace{2in} \\
& \quad \int_{t}^{t+\dd{t}}\left[ \f{\p  f_{m, n}}{\p s}
    + \sum_{i=1}^m g(X^i_s, A^i_s, s)\f{\p  f_{m, n}}{\p X^i_s} 
+ \sum_{j=1}^{2n}g(Y^j_s, B^{[(j+1)/2]}_s, s)
    \f{\p f_{m, n}}{\p Y^j_s}\right. \\[-6pt]
& \qquad\qquad\qquad\qquad\qquad\qquad\quad \left. 
+\f{1}{2}\sum_{i=1}^m\sigma^2(X^i_s, A^{i}_s, s)\f{\p^2 f_{m, n}}{(\partial{X^i_s})^2}
+\f{1}{2}\sum_{j=1}^{2n}\sigma^2(Y^j_s,B^{[(j+1)/2]}_s, s)\f{\p^2 f_{m, n}}{(\partial{Y^j_s})^2}\right]\dd s \\
& \qquad\qquad\quad  + \sum_{i=1}^m\int_{t}^{t+\dd{t}}
\!\!\sigma(X^i_s, A^{i}_s, s)\f{\p f_{m, n}}{\partial{X^i_s}} \dd W^i_s +
\sum_{j=1}^{2n}\int_{t}^{t+\dd{t}}\!\!\sigma(Y^j_s,B^{[(j+1)/2]}_s, s)
\f{\p f_{m, n}}{\partial{Y^j_s}}\dd \tilde{W}^j_s.
\label{Diff}
\end{aligned}
\end{equation}
%\end{multline}
%Note that by definition, $A_t^i=A_{t'}^i+t-t',
%B_t^j=B_{t'}^j+t-t'$.
%
After taking the expectation of Eq.~\eqref{Diff} we find

\begin{multline}
\mathbb{E}[f_{m, n}(\X^{(m)}_{t+\dd{t}},
\Y^{(2n)}_{t+\dd{t}}, t+\dd t ; \A^{(m)}_{t'}\!\!\!,\, \B^{(n)}_{t'})] -
\mathbb{E}[ f_{m, n}(\X^{(m)}_{t}\!\!\!,\,\Y^{(2n)}_{t}\!\!\!,\, t; \A^{(m)}_{t'}\!\!\!,\, 
\B^{(n)}_{t'})] =  \\
\mathbb{E}\left[\int_{t}^{t+\dd{t}}\!\!\!\!\dd s \left(\f{\p  f_{m, n}}{\p s} 
+ \sum_{i=1}^mg(X^i_s,A^i_s,s)\f{\p  f_{m, n}}{\p X^i_s}
+\sum_{j=1}^{2n}g(Y^j_s, B^{[(j+1)/2]}_s,s)\f{\p  f_{m, n}}{\p
  Y^j_s}\right.\right. \\
\hspace{-4mm} \left.\left. + \f{1}{2}\sum_{i=1}^m
\f{\p^2 f_{m, n}}{(\partial{X^i_s})^2}\sigma^2(X^i_s,
  A^{i}_s, s)+\f{1}{2}\sum_{j=1}^{2n}
\f{\p^2 f_{m, n}}{(\partial{Y^j_s})^2}\sigma^2(Y^j_s, B^{[(j+1)/2]}_s,s)\right)\right].
\label{ITODIFF}
\end{multline}
%
%
%For some $\X^{2n}=(X^1,...,X^m), X^i>0, \Y^{2n}=(Y^1, ..., Y^{2n}), Y^j>0$,
%
Specifically, we can take $f_{m, n}$ in Eq.~\eqref{ITODIFF}
as a distribution of the form
\begin{equation}
f_{m, n}(\X^{(m)}_t\!\!\!,\,\Y^{(2n)}_t\!\!\!,\,t; \A^{(m)}_{t'}\!\!\!,\,\B^{(n)}_{t'}) 
=  \prod_{i=1}^m
\delta(X^i-X^i_t) \prod_{j=1}^{2n}\delta(Y^j-Y^j_t) S_{1, m}(t;t',\A^{(m)}_{t'})
S_{2, n}(t;t',\B^{(m)}_{t'}),
\label{fdefinition}
\end{equation}
where $S_{1, m}$ and $S_{2, n}$ are \textbf{joint survival possibilities}

\begin{align}
S_{1, m}(t; t', \A^{(m)}) = &
\prod_{i=1}^m e^{-\int_{t'}^{t}\beta_{m, n}(A^i-t'+s,s)\textrm{ds}},
\quad\,\, S_{2, n}(t; t', \B^{(n)}) = \prod_{j=1}^n(e^{-\int_{t'}^{t}\beta_{m, n}(B^j-t'+s,s)\textrm{ds}})^2,
\label{JointSurv}
\end{align}
where the birth rate $\beta\equiv \beta_{m,n}$ can implicitly depend on the 
populations $m,n$. 

Next, we define $\hat{p}(\X_t^{(m)}\!\!\!\!,\, \Y_t^{(2n)}\!\!\!\!,\,
t\vert \X^{(m)}_{t'}\!\!\!,\, \Y_{t'}^{(2n)}\!\!\!\!,\,
\A_{t'}^{(m)}\!\!\!,\, \B_{t'}^{(n)})$ as the probability density of
$m$ singlets of volumes $\X^{(m)}_t$ and $n$ doublets of volumes
$\Y^{(2n)}_t$ at time $t$, conditioned on there being $m$ singlets of
volumes $\X^{(m)}_{t'}$ and ages $\A^{(m)}_{t'}$ and $n$ doublets with
volumes $\Y^{(2n)}_{t'}$ and ages $\B^{(2n)}_{t'}$ at time $t'$, and
that no cell division occurs during $[t', t]$.  The quantity
$\hat{p}(\X_t^{(m)}\!\!\!,\, \Y_t^{(2n)}\!\!\!\!,\, t\vert
\X^{(m)}_{t'}\!\!\!,\, \Y^{(2n)}_{t'}\!\!\!\!,\allowbreak \,
\A^{(m)}_{t'}\!\!\!,\, \B^{(n)}_{t'})S_{1, m}(t;t',\A^{(m)}_{t'})S_{2,
  n}(t;t',\B^{(m)}_{t'})$ is thus the probability measure that the
cell population at time $t$ contains $m$ singlets of size $\X_t^{(m)}$
and $n$ doublets of size $\Y_t^{(n)}$ with no cell division occurring
within $[t', t]$, conditioned on it containing $m$ singlets with
volumes $\X^{(m)}_{t'}$ and ages $\A^{(m)}_{t'}$ and $n$ doublets with
volumes $\Y^{(2n)}_{t'}$ and ages $\B^{(n)}_{t'}$ at $t'$.

After substitution of the $f_{m, n}$ defined in Eq.~\eqref{fdefinition}
into Eq.~\eqref{ITODIFF}, dividing by $\dd{t}$, and taking the
$\dd{t}\rightarrow 0$ limit, we obtain

\begin{equation}
\begin{aligned}
  \: & {\partial \over\partial t}\left(\hat{p}(\X^{(m)}\!\!,\, \Y^{(2n)}\!\!\!,\,
  t\vert \X^{(m)}_{t'}\!\!\!,\, \Y_{t'}^{(2n)}\!\!\!\!,\, \A_{t'}^{(m)}\!\!\!,\, 
\B_{t'}^{(n)}) S_{1,
    m}(t;t',\A_t^{(m)})S_{2, n}(t;t',\B_t^{(n)})\right) = \\  
  \: & \quad  \int_{\L^m}\!\!\dd{\X^{(m)}_t}\!\!\int_{\L^{2n}}\!\!\!\dd{\Y^{(2n)}_t}\, 
\hat{p}(\X_t^{(m)}\!\!\!,\, \Y_t^{(2n)}\!\!\!\!,\, t \vert \X^{(m)}_{t'}\!\!\!,\, 
\Y_{t'}^{(2n)}\!\!\!\!,\, \A_{t'}^{(m)}\!\!\!,\, \B_{t'}^{(n)})
\bigg[\f{\p f}{\p t} + \sum_{i=1}^{m} g(X^i_t, A^i_t, t)\f{\p f}{\p X^i_t} \\
\: & \qquad +\sum_{j=1}^{2n}g(Y^j_t, B^{[(j+1)/2]}_t, t)\f{\p f}{\p
      Y^j_t}+\f{1}{2}\sum_{i=1}^m\f{\p^2f}{\partial (X^i_t)^2}\sigma^2(X^i_t,A^{i}_t,t)
    +\f{1}{2}\sum_{j=1}^{2n}\f{\p^2f}{\partial (Y^j_t)^2}\sigma^2(Y^j_t,B^{[(j+1)/2]}_t,t) 
\bigg] \\
\: &  =-\Bigg[\bigg(\sum_{i=1}^{m}\beta_{m, n}(A_{t}^i, t)
+2\sum_{j=1}^{n}\beta_{m, n}(B_{t}^j, t)\bigg) \hat{p}_{m, n}
%S_{1, m}S_{2, n}
+\sum_{i=1}^m\f{\p (\hat{p}g(X^i_t, A^i_t, t))}{\p X^i_t} +
\sum_{j=1}^{2n}\f{\p (\hat{p}g(Y^j_t, B^{[(j+1)/2]}_t, t))}{\p Y^j_t}
 \\
% S_{1, m}S_{2, n} \\
\: &\hspace{4.8cm} -\f{1}{2}
\sum_{i=1}^{m}\f{\p^2 (\hat{p}\sigma^2(X^i_t, A^i_t, t))}{(\p X^i_t)^2}-
\f{1}{2} \sum_{j=1}^{2n}\f{\p^2 (\hat{p}\sigma^2(Y^j_t, B^j_t, t))}{(\p Y^j_t)^2}\Bigg]
S_{1, m}S_{2, n},
  \label{PHAT}
\end{aligned}
\end{equation}
where the last equality arises from integration by parts.

%%%%%%%%%%%%%%%%%%%%%%%%%%%%%%%%%%%%%%%%%%%%%%%%%%%%%%%%%%%%%%%%%%%%%%%%
\begin{comment}
%
Next, note that the quantity $\hat{p}(\X_t^m \Y_t^{2n}, {t}\vert
\X^m_{t'}, \Y^{2n}_{t'}, \A^m_{t'}, \B^{n}_{t'})S_{1,
  m}(t;t',\A^m_{t'})S_{2, n}(t;t',\B^m_{t'})$. Therefore, it satisfies
%
\small
\begin{equation}
\begin{aligned}
  &\f{\p (\hat{p}_{m, n}(\X_t^m
\Y_t^{2n}, {t}\vert \X^m_{t'}, \Y^{2n}_{t'}, \A^m_{t'}, 
\B^{n}_{t'})S_{1, m}(t;t',\A^m_{t'})S_{2, n}(t;t',\B^m_{t'}))}{\p t} 
+ \sum_{i=1}^m\f{\p (g(X^i_t, A^i_t, t)\hat{p}_{m, n}S_{1, m}S_{2, n})}{\p X_t^i}+ 
\sum_{j=1}^{2n}\f{\p (g(Y^j_t, B^j_t, t)\hat{p}_{m, n}S_{1, m}S_{2, n})}{\p Y_t^j} 
\\
&\:\quad\quad= - \bigg[\sum_{i=1}^m\beta_{m, n}(A_t^i, t)+
2\sum_{j=1}^n\beta__{m, n}(B_{t}^j, t)\bigg]\hat{p}_{m, n}S_{1, m}S_{2, n} 
+\f{1}{2}\sum_{i=1}^{m}\f{\p^2 (\sigma^2(X_t^i, A^i_t, t)
\hat{p}_{m, n}S_{1, m}S_{2, n})}{(\p X_t^i)^2}
+ \f{1}{2}\sum_{j=1}^{2n}\f{\p (\sigma^2(Y_t^j, B^j_t, t)
\hat{p}_{m, n}S_{1, m}S_{2, n})}{(\p Y_t^j)^2}.
\end{aligned}
\label{CondDiff}
\end{equation}
%
\end{comment}
%%%%%%%%%%%%%%%%%%%%%%%%%%%%%%%%%%%%%%%%%%%%%%%%%%%%%%%%%%%%%%%%%%%%%%%%

Finally, we derive the PDE satisfied by the unconditioned probability
density $p_{m,n}(\X^{(m)}_t\!\!\!,\, \Y^{(2n)}_t\!\!\!,\,
\A^{(m)}_t\!\!\!,\, \B^{(n)}_t\!\!\!,\, t)$ given $p_{m,
  n}(\X^{(m)}\!\!\!,\, \Y^{(2n)}\!\!\!\!,\,
\A^{(m)}\!\!\!,\,\B^{(n)}\!\!\!,\, t')$.  First, we note that if no
division has occurred in $[t', t]$ and $t-t'<\min\{A^{(m)}_t\!\!\!,\,
B^{(n)}_t\}$, a system at $t$ with $m$ singlets of volumes
$\X^{(m)}_t$ and ages $\A^{(m)}_t$ and $n$ doublets with volumes
$\Y^{(2n)}_t$ and ages $\B^{(n)}_t$ can result only from a system at
$t'$ with $m$ singlets with ages
$\A^{(m)}_{t'}\!\!=\A^{(m)}_{t}\!\!-(t-t')$ and $n$ doublets with ages
$\B^{(n)}_{t'}=\B^{(n)}_{t}\!\!- (t-t')$. Thus, we use the
Chapman-Kolmogorov relation between the two quantities
$\hat{p}(\X_t^{(m)}\!\!\!, \Y_t^{(2n)}\!\!\!,\,t \vert
\X^{(m)}_{t'}\!\!\!,\, \Y^{(2n)}_{t'}\!\!\!\!,\,
\A^{(m)}_{t'}\!\!\!,\,\B^{(n)}_{t'}) S_{1, m}(t;t',\A^{(m)}_{t'})S_{2,
  n}(t;t',\B^{(m)}_{t'})$ and $p_{m,n}$ to construct

%
%$\A^m_t>0, \B^n_t>0$.
%
%For $t>t'$, we wish to calculate $p_{m, n}(\X^m, \Y^{2n}, \A^m, \B^n,
%t)$ given $p_{m, n}(\X^m, \Y^{2n}, \A^m, \B^n, t')$.
%
%When $t-t'<\min\{A^m_t, B^n_t\}$, a system at $t$ with $m$ singlets of
%volumes $\X^m_t$ and ages $\A^m_t$ and $n$ doublets with volumes
%$\Y^{2n}_t$ and ages $\B^n_t$ can only result from a system at $t'$ of
%with $m$ singlets with age $\A^m_{t'}=\A^m_t-(t-t')$ and $n$ doublets
%with age $\B^n_{t'}=\B^n_t - (t-t')$ when there is no cell division
%during $[t', t]$. So

\begin{multline}
  p_{m, n}(\X^{(m)}_t\!\!\!,\, \Y^{(2n)}_t\!\!\!\!,\,
  \A^{(m)}_{t'}\!\!\!+t-t',\, \B^{(n)}_{t'}\!\!\!+t-t',\, t)   = 
  \int_{{\L^+}^{(m+2n)}}\!\!\!\!\hat{p}(\X_t^{(m)}\!,\,
\Y_t^{(2n)}\!\!\!, t\vert \X^{(m)}_{t'}\!\!\!,\, \Y^{(2n)}_{t'}\!\!\!\!,\, 
\A^{(m)}_{t'}\!\!\!,\B^{(n)}_{t'}) \\
\times S_{1, m}(t;t',\A^{(m)}_{t'})S_{2, n}(t;t',\B^{(m)}_{t'})
 p_{m, n}(\X^{(m)}_{t'}\!\!\!,\, \Y^{(2n)}_{t'}\!\!\!\!,\, \A^{(m)}_{t'}\!\!\!,\,
\B^{(n)}_{t'}\!\!\!,\, t')\dd{\X^{(m)}_{t'}}\dd{\Y^{(2n)}_{t'}}\!\!.
\label{Uncond}
\end{multline}
Assuming that $p_{m, n}$ is continuous and differentiable, and the
integration is interchangeable with differentiation in
Eq.~\eqref{Uncond}, we take derivatives with respect to all variables
$t, X^i, Y^j, A^i, B^j$ to obtain
\begin{equation}
\begin{aligned}
&\f{\p p_{m, n}}{\p t}
%(\X^{(m)}_t\!\!\!,\, \Y^{(2n)}_t\!\!\!\!,\, \A^{(m)}_t\!\!\!,\,\B^{(n)}_t\!\!\!,\, t)
+\sum_{i=1}^m\f{\p (g(X^i_t, A^i_t, t)p_{m, n})}{\p X^i_t}
+\sum_{j=1}^{2n}\f{\p (g(Y^j_t, B^j_t, t)p_{m, n})}{\p Y^j_t}
+\sum_{i=1}^m\f{\p p_{m, n}}{\p A^i_t}+\sum_{j=1}^{n}\f{\p p_{m, n}}{\p B^j_t} 
=  \\
& - \bigg(\sum_{i=1}^m\beta_{m, n}(A^i_t, t)+2\sum_{j=1}^n\beta_{m, n}(B^j_t, t)\bigg)p_{m, n} 
+ \f{1}{2}\sum_{i=1}^{m}\f{\p^2 (\sigma^2(X^i_t, A^i_t, t)p_{m, n})}{(\p X^i_t)^2} 
+ \f{1}{2}\sum_{j=1}^{2n}\f{\p (\sigma^2(Y^j_t, B^j_t, t)p_{m, n})}{(\p Y^j_t)^2},
%
%
%=\int_{{\L}^{(m+2n)}}\!\dd{\X^m_{t'}}\dd{\Y^{2n}_{t'}}\, 
%p_{m, n}(\X^m_{t'}, \Y^{2n}_{t'}, \A^m_{t'}, \B^n_{t'}, t')
%
%\f{\p (\hat{p}(\X_t^m,\Y_t^{2n}, {t}\vert \X^m_{t'}, \Y^{2n}_{t'}, \A^m_{t'}, 
%\B^{n}_{t'})S_{1, m}(t;t',\A^m_{t'})S_{2, n}(t;t',\B^m_{t'}))}{\p t}\\
%
\label{TransportSDE}
\end{aligned}
\end{equation}
where $p_{m, n}\equiv p_{m,n}(\X^{(m)}_t\!\!\!,\, \Y^{(2n)}_t\!\!\!,\,
\A^{(m)}_t\!\!\!,\,\B^{(n)}_t, t)$.  Hereafter, we will omit the
subscript $t$ for notational simplicity.  To facilitate further
analysis, we define a symmetrized density $\rho_{m, n}$ that is
symmetric to the interchange of variables:

\begin{equation}
  \rho_{m, n}(\X^{m}\!\!,\, \Y^{2n}\!\!\!,\, \A^{m}\!\!\!,\,\B^{n}\!\!,\,t)
  = \f{1}{2^{n}m!n!} 
  \sum_{\pi^{2n}} p_{m, n}(\X^{(m)}\!\!\!,\, \pi^{2n}(\Y^{(2n)}), \A^{(m)}\!\!\!,\,
  \B^{(n)}\!\!\!,\, t)
\label{rhodefinition}
\end{equation}
where $\A^{(m)}=(A^{\xi_a(1)},\ldots, A^{\xi_a(m)})$,
$\B^{(n)}=(B^{\xi_b(1)},\ldots, B^{\xi_b(m)})$ are ordered ages,
$\X^{(m)}=(X^{\xi_a(1)},\ldots,X^{\xi_a(m)})$, \\ $Y^{(2n)} =
(Y^{2\xi_b(1)-1},\ldots, Y^{2\xi_b(n)})$ are the corresponding sizes,
and $\pi^{2n}$ is some permutation $\L^{2n}\rightarrow\L^{2n}$ such
that $\pi^{2n}(Y^{2i}), \pi^{2n}(Y^{2i-1})\in \{Y^{2i-1}, Y^{2i}\},
\pi^{2n}(Y^{2i})\neq \pi^{2n}(Y^{2i-1}), i=1,...,n$, \textit{i.e.},
$\pi^{2n}$ can interchange the sizes of two cells in a doublet.
Therefore, there are $2^n$ total permutations
$\pi^{2n}$. $\xi_a(1), ..., \xi_a(m)$ is a rearrangement such that
$A^{\xi_a(1)}\geq A^{\xi_a(2)}\geq...\geq A^{\xi_a(m)}$ and $\xi_b(1),
..., \xi_b(n)$ is a rearrangement such that $B^{\xi_b(1)}\geq
B^{\xi_b(2)}\geq...\geq B^{\xi_b(m)}$. Defining such a $\rho_{m, n}$
allows us to remove the restriction that the ages must be presented in
a descending order.  Moreover, changing the order of two cells within
in a doublet will not affect the value of $\rho_{m, n}$. Definite
integrals over $\rho_{m,n}$ are then related to those over $p_{m, n}$
via

\begin{equation}
\begin{aligned}
&\int
%_{{\L}^{(m+2n+m+n)}}
\!\dd\X^m\dd\Y^{2n}\dd\A^m\dd\B^{n}\rho_{m, n}(\X^m, \Y^{2n}, \A^m, \B^{n},t) = 
\int
%_{{\L}^{(m+2n)}}
\!\dd\X^{(m)}\dd\Y^{(2n)} \int_{\L}\!\dd{A^{\xi_a(1)}}...\\
&\:\qquad...\int_0^{A^{\xi_a(m-1)}}\!\!\!\!\!\!\!\dd{A}^{\xi_a(m)}
\int_{\L}\!\dd{B^{\xi_b(1)}}...
\int_0^{B^{\xi_b(n-1)}}\!\!\!\!\!\!\!\!\dd{B}^{\xi_b(n)}\, 
p_{m, n}(\X^{(m)}\!\!\!,\, \Y^{(2n)}\!\!\!,\, \A^{(m)}\!\!\!,\, \B^{(n)}\!\!\!,\, t),
\end{aligned}
\end{equation}
so $\rho_{m,n}$ is also a probability density distribution if $p_{m,
  n}$ is. Furthermore, the differential equation satisfied by
$\rho_{m, n}$ for $\A^m\!\!,\,\B^n>0$ is the same as the differential
equation satisfied by $p_{m, n}$
\begin{equation}
\begin{aligned}
&\f{\p \rho_{m, n}}{\p t}+ \sum_{i=1}^m\f{\p \rho_{m, n}}{\p A^i} +  
\sum_{j=1}^{n}\f{\p \rho_{m, n}}{\p B^j} + 
\sum_{i=1}^m\f{\p (\rho_{m, n}g(X^i, A^i, t))}{\p X^i} +
\sum_{j=1}^{2n}\f{\p (\rho_{m, n}g(Y^j, B^{[\f{j+1}{2}]}, t))}{\p Y^j} 
=  \\
&\hspace{3cm} -\bigg(\sum_{i=1}^m\beta_{m, n}(A^i, t)
+2\sum_{j=1}^n\beta_{m, n}(B^j, t)\bigg)\rho_{m, n} \\
& \hspace{5cm} + \f{1}{2}\sum_{i=1}^{m}
\f{\p^2 (\sigma^2(X^i, A^i, t)\rho_{m, n})}{(\p X^i)^2} +
\f{1}{2}\sum_{j=1}^{2n}\f{\p^2 (\sigma^2(Y^j, B^{[\f{j+1}{2}]}, t)
\rho_{m, n})}{(\p Y^j)^2}.
\label{RHO}
\end{aligned}
\end{equation}

\subsection{Boundary Conditions}
We now specify appropriate boundary conditions for $\rho_{m, n}$ that
represent the birth of new cells with age zero. By using ordered ages,
it is easy to derive the corresponding boundary conditions for $p_{m,
  n}$ defined in Eq.~\eqref{Uncond}, which we omitted here, but which
are nonzero if $B^n=0$ and zero if any entry in $\X^m, \Y^{2n}, \A^m,
\B^{k<n}$ is zero.  The boundary consitions for $\rho_{m, n}$ are then
derived from the boundary conditions for $p_{m, n}$. Homogeneous
boundary conditions also arise at any $X^i=0, \infty$ or $Y^j=0,
\infty$ indicating that no cell can have 0 or infinite size.
%
%The boundary condition if a $B^j=0$ comes from division of existing
%cells and birth of new cells.
%
If one cell divides at time $t$ in a system of $m$ singlets and $n$
doublets, the system could either convert to $m-1$ singlets and $n+1$
doublets when this dividing cell is a singlet, or $m+1$ singlets and $n$
doublets when the dividing cell is one cell in a doublet.
A simpler but similar discussion of boundary conditions for the
``timer" model which has no size dependence has been discussed
\cite{chou2016hierarchical_PRE,chou2016hierarchical}. Hereafter, we
use the notation $\X^m_{-i}= (X^1,X^2,...,X^{i-1}, X^{i+1},...,X^m)$,
$\A^m_{-i}= (A^1,A^2,...,A^{i-1}, A^{i+1},...,A^m)$ to describe
vectors of one lower dimension in which element $i$ is removed.  The
boundary conditions are described by

\begin{equation}
\rho_{m, n} = 0 \,\,  \left\{\begin{array}{l}
\mbox{if any element in}~\{\X^m, \Y^{2n}\} = 0,\infty, \\
\mbox{or more than one  element in}~\A^m=0, \\
\mbox{or more than one  element in}~\B^n = 0,\end{array}\right.
\label{BCrho00}
\end{equation}
and
\begin{equation}
\begin{aligned}
& \rho_{m, n}(\X^m\!\!\!,\, \Y^{2n}[Y^{2j-1}\!=y_1,Y^{2j}\!=y_2],\A^{m}\!\!,\, \B^{n}[B^j=0], t) = \\
& \f{m+1}{n}\!\int_{0}^{\infty}\!\!\!\,\tilde{\beta}_{m+1, n-1}( y_1+y_2, y_1,s, t)
\rho_{m+1, n-1}( \X^{m+1}[X^{m+1}=y_1+y_2], \Y^{n-1}, \A^{m+1}[A^{m+1}=s], \B^{n-1}, t)\dd{s} \\
& \quad +\f{2}{m}\!\sum_{i=1}^{m}\tilde{\beta}_{m-1, n}(y_1+y_2, y_1, A^i, t)
\rho_{m-1, n}(\X^m_{-i},\A^m_{-i}, \B^n[B^n=A^i], \Y^{2n}[Y^{2n-1}={X}^i, Y^{2n}=y_1+y_2], t),
\label{BCrho}
\end{aligned}
\end{equation}
where $\tilde{\beta}_{m, n}(x, z, a, t)\dd{z}$ is the differential
rate that, in a population of $m$ singlets and $n$ doublets, a cell of
volume $x$ and age $a$ divides into one
  cell with volume $\in [z, z+\dd z]$. From volume conservation,
  $\tilde{\beta}_{m, n}(x, z, a, t)=\tilde{\beta}_{m, n}(x, x-z, a,
  t)$, and if we assume the form $\tilde{\beta}_{m, n}(x, z, a, t) =
  h(z/x)\beta_{m,n}(a,t)/x$ \cite{Xia2020}, $\int_0^x\tilde{\beta}_{m,
    n}(x, z, a, t)\dd{z}=\beta_{m, n}(a, t)$ is independent of size
  $x$ as we have assumed. The notation $\X^{m+1}[X^i=x]$ indicates
  that the $i^{\textrm{th}}$ component in $\X^{m+1}$ is $x$, with
  similar definitions for $\Y^{2n}[Y^j=y], \A^{m}[A^i=a],
  \B^n[B^j=b]$. The zero-valued conditions in Eq.~\eqref{BCrho00}
  enforces that no cell can have zero or infinitely large volume and
  that no more than one cell can divide at the same time (continuous
  time assumption).  The first term on the RHS of Eq.~\eqref{BCrho}
  results from the division of a singlet while the second term results
  from the division of one cell in a doublet, leaving a singlet and
  giving rise to a new doublet.  Finally, in the Appendix, we
  explicitly demonstrate that probability conservation is preserved
  under these boundary conditions.

\section{Hierarchies and moment equations}
\label{meanfield}
In this section, we will assume that $\tilde{\beta}$ and $\beta$ are
independent of the population sizes $m, n$. Under this assumption, we
are able to derive lower-dimensional (\textit{e.g.}, marginalized)
projections of our kinetic theory (Eq.~\eqref{RHO}) by averaging over
a variable number of cell sizes:

\begin{equation}
\begin{aligned}
&\rho_{m, n}^{(h, k, \ell)}
(\X^h\!\!,\, \A^{h}\!\!,\,\Y_{\rm e}^{2k+2\ell}\!\!\!\!\!,\,\, \B^{k+\ell}\!\!\!,\, t) 
= \int_{\L}\dd \X^{h+1:m}
\dd{\Y_{\rm o}}^{2k+2\ell+1:2n}\dd{\A}^{h+1:m}\dd{\B}^{k+\ell+1:n}\,
\rho_{m, n},
%(\X^m\!\!,\, \Y^{2n}\!\!\!,\, \A^m\!\!,\, \B^{n}\!\!,\, t),
\end{aligned}
\end{equation}
where $\rho_{m, n}\equiv \rho_{m, n}(\X^m\!\!,\, \Y^{2n}\!\!\!,\,
\A^m\!\!,\, \B^{n}\!\!,\, t)$, $\L \equiv
\L^{(m-h)+(2n-k-2\ell)+(m-h)+(n-k)}$, and we define the notation
$\X^{h+1:m}\!\coloneqq(X^{h+1},...,X^m), \Y_{\rm
  o}^{2k+2\ell+1:2n}\!\coloneqq(Y^1, Y^3,...,Y^{2k-1},
Y^{2k+2\ell+1},...,Y^{2n})$, $\A^{h+1:m}\!\coloneqq(A^{h+1},...,A^m)$,
$\B^{k+\ell+1:n}\!\coloneqq(B^{k+\ell+1},\\
...,B^n)$ and $\Y_{\rm
  e}^{2k+2\ell}\!\coloneqq (Y^2, Y^4,...,Y^{2k},
Y^{2k+1},Y^{2k+2},...,Y^{2k+2\ell})$.  The marginalized densities
require three indices to describe because although the size $\X^m$ and
age $\A^m$ have a one-to-one correspondence for singlets, the twins,
while carrying the same age, almost surely have different sizes due to
asymmetric division and independent growth fluctuations immediately
after birth. Thus, the number of ways to exit and enter each state
depends on which types of cells are ``integrated over".  By
marginalizing over Eq.~\eqref{RHO}, we find the kinetic equation
satisfied by $\rho_{m, n}^{(h, k, \ell)}$ (in the remaining space
$\X^h\!, \A^h\!, \Y_{\rm e}^{2k+2\ell}\!\!\!,\, \B^k>0$) becomes
\begin{equation}
\begin{aligned}
&\f{\p \rho_{m, n}^{(h, k, \ell)}(\X^h\!\!,\, \A^{h}\!\!,\,
\Y_{\rm e}^{2k+2l}\!\!,\, \B^{k+\ell}\!\!,\, t)}{\p t} 
+\sum_{i=1}^{h}\f{\p \rho_{m, n}^{(h, k, l)}}{\p A^i} + \sum_{j=1}^{k+\ell}
\f{\p \rho_{m, n}^{(h, k, \ell)}}{\p B^j}
+ \sum_{i=1}^{h}\f{\p (g(X^i, A^i, t)\rho_{m, n}^{(h, k, \ell)})}{\p X^i}  \\
&\: \hspace{8mm} + \sum_{j=1}^{k} \f{\p (g(Y^{2j}, A^j, t)\rho_{m, n}^{(h, k, \ell)})}{\p Y^{2j}} 
+ \sum_{j=1}^{2\ell} \f{\p (g(Y^{2k+j}\!\!\!,\, A^j\!, t)
\rho_{m, n}^{(h, k, \ell)})}{\p Y^{2k+j}}- 
\f{1}{2}\sum_{i=1}^h\f{\p^2 (\sigma^2(X^i\!\!,\, A^i\!\!,\, t)
\rho_{m, n}^{(h, k, \ell)})}{(\p X^i)^2} \\
&\: \hspace{3.2cm} -\f{1}{2}\sum_{j=1}^{k}\f{\p^2 (\sigma^2(Y^{2j}\!\!,\, B^{[\f{j+1}{2}]}\!, t)
\rho_{m, n}^{(h, k,\ell)})}{(\p Y^{2j})^2}
-\f{1}{2}\sum_{j=1}^{2\ell}\f{\p^2 (\sigma^2(Y^{2k+j}\!\!\!,\, B^{k+[\f{j+1}{2}]}\!, t)
\rho_{m, n}^{(h, k, \ell)})}{(\p Y^{2k+j})^2}\\
&\: = -\sum_{i=1}^h\beta(A^i\!, t)\rho_{m, n}^{(h, k, \ell)}(\X^h\!\!,\, 
\Y_{\rm e}^{2k+2\ell}\!\!\!,\, \A^{h}\!\!,\, \B^{k+\ell}\!\!\!,\, t)
- \sum_{j=1}^{k+\ell}2\beta(B^j\!, t)
\rho_{m, n}^{(h, k, \ell)}(\X^h\!\!,\, \Y_{\rm e}^{2k+2\ell}\!\!\!,\, \A^{h}\!\!,\, 
\B^{k+\ell}\!\!\!,\, t)\\
&\: \quad -(m-h)\int_{\L^2}\dd{X^{h+1}}\dd{A^{h+1}}\, \beta(A^{h+1}\!\!\!,\, t)
\rho_{m, n}^{(h+1, k, \ell)}(\X^{h+1}\!\!\!,\, \Y_{\rm e}^{2k+2\ell}\!\!\!\!,\,
\A^{h+1}\!\!\!,\, \B^{k+\ell}\!\!\!,\, t) \\
&\: \quad - 2(n-k-\ell)\int_{\L^2}\dd{Y^{2k+2}}\dd{B^{k+1}}\,
\beta(B^{k+1}\!\!\!,\, t)
\rho_{m, n}^{(h, k+1, \ell)}(\X^h\!\!,\, \Y_{\rm e}^{2k+2\ell+2}\!\!\!\!,\, \A^{h}\!\!,\, \B^{k+\ell+1}\!\!\!,\, t)\\
&\: \quad +\f{(n-k-\ell)(m+1)}{n}\int_{\L^2}\dd{X^{h+1}}\dd{A^{h+1}}\beta(A^{h+1}, t)\rho_{m+1, n-1}^{(h+1, k, \ell)}
(\X^{h+1}\!\!\!,\, \Y_{\rm e}^{2k+2\ell}\!\!\!\!,\, \A^{h+1}\!\!\!,\, \B^{k+\ell}\!\!\!,\, t)  \\
&\:\quad+\f{2(n-k-\ell)(m-h)}{m}\!\int_{\L^2}\!\!
\dd{Y^{2k+2}}\dd{B^{k+1}}\,\beta(B^{k+1}\!\!\!,\, t)
\rho_{m-1, n}^{(h, k+1, \ell)}(\X^h\!\!,\,\Y_{\rm e}^{2k+2\ell+2}\!\!\!\!,\,\A^{h}\!\!,\, 
\B^{k+\ell+1}\!\!\!\!,\, t)\\
&\:\quad+\f{2(n-k-\ell)}{m}\sum_{i=1}^{h} \beta(A^i, t)  
\rho_{m-1, n}^{(h-1, k+1, \ell)}(\X^h_{-i}, \Y_{\rm e}^{2k+2+2\ell}[Y^{2k+2}=X^i], 
\A^{h}_{-i},\, \B^{k+\ell+1}[B^{k+1}=A^i], t),\label{rho}
\end{aligned}
\end{equation}
\normalsize
and the associated boundary conditions become
\begin{equation}
\begin{aligned}
&\rho_{m, n}^{(h, k,\ell)}(\X^h\!\!,\, \Y_{\rm e}^{2k+2\ell}[Y^{2k}\!\!=y], \A^{h}\!\!\!,\, 
\B^{k+\ell}[B^k\!=0], t) = \\
&\:\quad\f{m+1}{n}\!\int_{\L^2}\!\!\dd{A^{h+1}}
\dd{s}\,\tilde{\beta}(s+y, y, A^{h+1}, t)
\rho_{m+1, n-1}^{(h+1, k-1, \ell)}(\X^{h+1}[X^{h+1}\!\!=  s+y], 
\Y_{\rm e}^{2k+2\ell-2}\!\!\!\!,\,\A^{h+1}\!\!\!,\, \B^{k+\ell-1}\!\!\!,\, t) \\
&\:\quad+\f{2(m-h)}{m}\int_{\L^2}\!\!\dd{B^{k}}\dd{s}\,
\tilde{\beta}(s+y, y, B^{k}\!\!,\, t)
\rho_{m-1, n}^{(h, k, \ell)}(\X^h\!\!,\, \Y_{\rm e}^{2k}[Y^{2k}\!\! = s+y], \A^{h}\!\!,\, \B^{k+\ell}\!\!\!,\, t)\\
&\:\quad+\f{2}{m}\sum_{i=1}^{h}\int_{\L}\!\!
\dd{s}\,
\tilde{\beta}(s+y, y, A^i\!, t)  \rho_{m-1, n}^{(h-1, k-1, \ell+1)}(\X^h_{-i}, 
\Y_{\rm e}^{2k+2\ell}[Y^{2k+2\ell-1}\!\!= s+y, Y^{2k+2\ell}\!\!=X^i],...\\[-5pt]
&\hspace{10.6cm}...\A^{h}_{-i}, \B^{k+\ell}[B^{k}\!= A^i], t),
\label{rhobc1}
\end{aligned}
\end{equation}

\begin{equation}
\begin{aligned}
&\rho_{m, n}^{(h, k, \ell)}(\X^h\!\!,\, \Y_{\rm e}^{2k+2\ell}[Y^{2k+2\ell-1}\!\!=y_1, 
Y^{2k+2\ell}\!\!\!=y_2], \A^{h}\!\!,\, \B^{k+\ell}[B^{k+\ell}\!\!=0], t) = \\
&\:\quad\f{m+1}{n}\!\int_{\L}\!\!\dd{A^{h+1}}\,\tilde{\beta}(y_1+y_2, y_1, A^{h+1}\!\!\!,\, t)
\rho_{m+1, n-1}^{(h+1, k, \ell-1)}(\X^{h+1}[X^{h+1}\!\!\!= y_1+y_2], 
\Y_{\rm e}^{2k+2\ell-2}\!\!\!,\,\A^{h+1}\!\!\!,\, \B^{k+\ell-1}\!\!\!,\, t) \\
&\:\quad+\f{2(m-h)}{m}\!\int_{\L}\!\!\dd{B^{k+1}}
\, \tilde{\beta}(y_1+y_2, y_1, B^{k+1}\!\!\!,\, t)
\rho_{m-1, n}^{(h, k+1, \ell-1)}(\X^h\!\!, \Y_{\rm e}^{2k+2\ell}[Y^{2k+2}\!\!= y_1+y_2], 
\A^{h}\!\!,\, \B^{k+\ell}\!\!\!,\, t) \\
&\:\quad+\f{2}{m}\sum_{i=1}^{h}\tilde{\beta}(y_1+y_2, y_1, A^i\!, t) 
\rho_{m-1, n}^{(h-1, k, \ell)}
(\X^h_{-i}, \Y_{\rm e}^{2k+2\ell}[Y^{2k+2\ell-1}\!\!= y_1+y_2, Y^{2k+2\ell}\!\!=X^i],...\\[-4pt]
& \hspace{10.4cm}...\A^{h}_{-i}, \B^{k+\ell}[B^{k+\ell}\!= A^i], t),
\label{rhobc}
\end{aligned}
\end{equation}
and 

\begin{align}
&\rho_{m, n}^{(h, k, \ell)}(\X^h[X^i=0], \Y_{\rm e}^{2k+2\ell}\!\!\!\!,\, 
\A^{h}\!\!,\, \B^{k+\ell}\!\!\!,\,t)=
\rho_{m, n}^{(h, k, \ell)}(\X^h[X^i=\infty], \Y_{\rm e}^{2k+2\ell}\!\!\!,\,
\A^{h}\!\!,\, \B^{k+\ell}\!\!\!,\, t)=0, \,\,\, i=1,2,...,h,\\
&\rho_{m, n}^{(h, k, \ell)}(\X^h\!\!,\, \Y_{\rm e}^{2k+2\ell}[Y^j\!=0],
\A^{h}\!\!,\, \B^{k+\ell}\!\!\!\!,\, t)=
\rho_{m, n}^{(h, k)}(\X^h\!\!,\,\Y_{\rm e}^{2k+2\ell}[Y^j\!=\infty], \A^{h}\!\!,\, 
\B^{k+\ell}\!\!\!,\, t)=0, \nonumber \\
& \:\hspace{9.9cm}\quad j=2,4,...,2k,2k+1,...,2k+2\ell, \\
&\:\qquad\rho_{m, n}^{(h, k, \ell)}(\X^h\!\!,\, \Y_{\rm e}^{2k+2\ell}\!\!\!\!,\, \A^{h}[A^i=0], \B^{k}, t)=0, \hspace{1.75cm}  i=1,2,...,h,\\
&\:\qquad \rho_{m, n}^{(h, k, \ell)}(\X^h\!\!,\, 
\Y_{\rm e}^{2k+2\ell}\!\!\!\!,\, \A^{h}\!\!,\, \B^{k+\ell}\!\!\!,\, t)=0, 
\hspace{2.8cm} \textrm{if two or more entries in $\B^{k+\ell}$ are 0}.
\label{rhobc0}
\end{align}
The first two terms on the RHS of Eq.~\eqref{rho} represent the
division of a singlet/doublet in the current system whose age is
specified; the third and fourth terms on the RHS stand describe the
division of a singlet and one cell of a doublet, respectively, whose
age is not specified; the fifth term results from the division of a
singlet, whose age and volume are unspecified, that induces the state
transition $(m+1, n-1) \to (m, n)$. The sixth term arises from
division of one cell of a doublet that coverts the system from $(m-1,
n)$ to $(m,n)$.  Finally, the last term represents the division of one
cell in a doublet whose age is $A^i, 1\leq i \leq h$ and its
undividing twin has size $X^i$.  In Eq.~\eqref{rhobc1}
and~\eqref{rhobc}, the first term on their RHSs represent the division
of a singlet, and the second term on their RHSs describe the division
of one cell in a doublet, giving rise to a newborn doublet and leaving
a singlet whose volume and age are integrated over. The last terms in
the boundary conditions in Eq.~\eqref{rhobc1}
and~\eqref{rhobc} result from the division of a cell in a
doublet, resulting in a newborn doublet and leaving a singlet whose
volume and age are $X^{i}\in \X^h$ and $A^{i}\in \A^h$, respectively.

Our kinetic equations subsume all hierarchical equations for $\rho_{m,
  n}^{(h, k,\ell)}$. First, we consider the lowest order equations
$(h=k=\ell=0)$ and the physical quantities that can easily be
constructed such as the total number $N=m+2n$.  The total expected
cell population can be expressed as

\begin{equation}
\mathds{E}[N(t)] = \sum_{m=0}^{\infty}\sum_{n=0}^{\infty}(m+2n)\rho_{m, n}^{(0, 0, 0)},
\end{equation}
which satisfies

\begin{align}
\f{\dd \mathds{E}[N(t)]}{\dd t} & = 
\sum_{m=0}^{\infty}\sum_{n=0}^{\infty}\left[ m\!\int_{\L^2}\!\dd{X^1}\dd{A^1}\,
\beta(A^1\!\!,\, t)\rho_{m, n}^{(1, 0, 0)}(\X^1\!\!,\,\A^1\!\!,\, t)+2n\!\int_{\L^2}\!
\dd{Y^2}\dd{B^1}\,\beta(B^1\!\!,\, t)
\rho_{m, n}^{(0, 1, 0)}(\Y^2_{\rm e},\B^1\!\!,\, t)\right] \nonumber\\
& =\int_{\L^2}\!\dd{x}\,\dd{a}\, \beta(a, t)n^{(1, 0)}(x, a, t)
\label{Nequation}
\end{align}
and involves the higher-dimensional densities $\rho_{m, n}^{(1, 0,
  0)}$ and $\rho_{m, n}^{(0,1,0)}$. The differential equation for
$\mathds{E}[N(t)]$ does not involve a any boundary condition, but it
is not closed because it depends on $n^{(1, 0)}$.

% if we take all three subindices $h, k, \ell$. For simplicity, we do
% not include further discussions here.

Higher dimensional total number-density functions 
$n^{(k, \ell)}(\x^k,\y^{2\ell}, \a^k, \b^{\ell}, t)$ can also be
generally defined:

%\scriptsize
\begin{equation}
\begin{aligned}
&n^{(k, \ell)} = \sum_{m=0}^{\infty}\sum_{n=0}^{\infty}\sum_{r=0}^k
\sum_{\xi^{(0, r)}\in S_k}
%
%_{\xi^{(0, r)}\in\Omega_r, \xi^{(r, k-r)}= S_k\backslash \xi^{(0, r)}}
%
2^{k+\ell-r}(m)_r(n)_{k+\ell-r}\,
\rho_{m, n}^{(r, k-r, \ell)}(\X^r[X^i=x^{\xi^{(0, r)}(i)}],... \\
& \qquad\qquad ...,\Y^{2(k-r)+2\ell}_{\rm e}[Y^{2j}=x^{\xi^{(r, k-r)}(j)}, Y^{2(k-r)+p}=y^p],
\A^r[A^i=a^{\xi^{(0, r)}(i)}],... \\
&\qquad\qquad ...,\B^{k-r+\ell}[B^j=a^{\xi^{(r, k-r)}(j)}, 
B^{k-r+[\f{p+1}{2}]}=b^{[\f{p+1}{2}]}], t), 
\quad\, 1\leq i\leq r, 1\leq j \leq k-r, 1\leq p \leq 2\ell 
\label{quantity}
\end{aligned}
\end{equation}
where $\x^k\!\coloneqq (x^1,...,x^k), \y^{2\ell}\!\!\coloneqq
(y^1,...,y^{2\ell}), \a^k\!\coloneqq (a^1,...,a^k),
\b^{\ell}\!\coloneqq (b^1,...,b^{\ell})$, $(m)_{r} = m!/(m-r)!$ is the
falling factorial, $S_k=\{1, 2, ..., k\}$.  The sum
$\sum\limits_{\xi^{(0, r)}\in S_k}$ includes summing over all elements
$\xi^{(0, r)}\in\Omega_r$, the set that contains all possible choices
of choosing $r$ elements in $S_k$, and $\xi^{(r, k-r)}\!\coloneqq
(\xi(r+1), \xi(r+2),...\xi(k))=S_k\backslash\xi^{(0, r)}$.  We require
$\xi^{(0, r)}(i) < \xi^{(0, r)}(j), \xi^{(r, k-r)}(i) < \xi^{(r,
  k-r)}(j),\, \forall i<j$ and $r\leq m, k-r\leq n$ in
Eq.~\eqref{quantity}.  With a $\beta$ independent of $m,n$, the PDE
satisfied by $n^{(k, \ell)}(\x^k, \y^{2\ell}, \a^k, \b^{\ell}, t)$ is
\begin{align}
&\f{\p n^{(k, \ell)}}{\p t}
+ \sum_{i=1}^k\f{\p n^{(k, \ell)}}{\p a^i} + 
\sum_{j=1}^{\ell}\f{\p n^{(k, \ell)}}{\p b^j}
+ \sum_{i=1}^k\f{\p (n^{(k, \ell)}g(x^i, a^i, t))}{\p x^i} + 
\sum_{j=1}^{2\ell}\f{\p (n^{(k, \ell)}g(y^j, b^{[\f{j+1}{2}]}, t))}{\p y^j} =  \nonumber\\
&\,\,\, -\bigg(\sum_{i=1}^k \beta(a^i, t)n^{(k, \ell)}
+\sum_{j=1}^{\ell} 2\beta(b^j, t)\bigg)n^{(k, \ell)}+
\f{1}{2}\sum_{i=1}^k\f{\p^2 (n^{(k, \ell)}\sigma^2(x^i, a^i, t))}{(\p x^i)^2}+
\f{1}{2}\sum_{j=1}^{2\ell}\f{\p^2 (n^{(k, \ell)}\sigma^2(y^j, b^{[\f{j+1}{2}]}, t))}
{(\p y^j)^2},
\label{PDEhighdim}
\end{align}
along with the boundary conditions
%\tiny

\begin{equation}
\begin{aligned}
& n^{(k, \ell)}(\x^k[x^v=x], \a^k[a^v=0],\y^{2\ell}, \b^{\ell}, t) =
\sum_{m=0}^{\infty}\sum_{n=0}^{\infty}\sum_{r=0}^{k-1}\sum_{\xi^{(0, r)}\in S_k^{-v}}
%
%_{\xi^{(0, r)}\in\Omega_r^{-v}, \xi^{(r, k-r)}= S_k^{-v}
%\backslash \xi^{(0, r)}, \xi^{(r, k-r)}(k-r)=v}
%
2^{\ell+k-r}(m)_r(n)_{k+\ell-r}\times \\
&\:\qquad\qquad\quad \rho_{m, n}^{(r, k-r, \ell)}(\X^r[X^i=x^{\xi^{(0, r)}(i)}],\Y_{\rm e}^{2k-2r+2\ell}[Y^{2j}=x^{\xi^{(r, k-r)}(j)}, Y^{2k+p}=y^p],... \\
&\hspace{3.9cm} ...,\A^r[A^i=a^{\xi^{(0, r)}(i)}],\B^{\ell+k-r}[B^j=a^{\xi^{(r, k-r)}(j)}, 
B^{k-r+[\f{p+1}{2}]}=b^{[\f{p+1}{2}]}], t) \\
&\: \quad =2\int_{\L^2}\dd{s}\dd{a} \,\tilde{\beta}(x+s,x, a, t)
n^{(k, \ell)}(\x^k[x^k=x+s],\y^{2\ell}, \a^k[a^k=a],\b^{\ell}, t) \\
&\:\qquad\qquad + 2\sum_{u=1,\neq v}^k\int_{\L}\dd{s}\,\tilde{\beta}(x+s, x, a^u, t)
n^{(k-2, \ell+1)}(\x^{k}_{-u, -v}, \a^{k}_{-u, -v},...\\
& \: \hspace{5cm}...,\y^{2\ell+2}[y^{2\ell+1}=x^u, y^{2\ell+2}=s+x],\b^{\ell+1}[b^{\ell+1}=a^u], t)
\label{BChighdim}
\end{aligned}
\end{equation}
\begin{equation}
\begin{aligned}
& n^{(k, \ell)}(\x^k, \y^{2\ell}[y^{2v-1}=y_1, y^{2v}=y_2],\a^k, \b^{\ell}[b^v=0], t)= 
 \sum_{m=0}^{\infty}\sum_{n=0}^{\infty}\sum_{r=0}^{k}\sum_{\xi^{(0, r)}\in S_k}
2^{\ell+k-r}(m)_r(n)_{k+\ell-r}\times\\
&\:\qquad\quad\rho_{m, n}^{(r, k-r, \ell)}(\X^r[X^i=x^{\xi^{(0, r)}(i)}],\Y_{\rm e}^{2\ell+k-r}[Y^{2j}=x^{\xi^{(r, k-r)}(j)}, Y^{2k+q}=y^q],... \\
&\:\hspace{4.2cm}...,\A^r[A^i=a^{\xi^{(0, r)}(i)}],\B^{\ell+k-r}[B^j=a^{\xi^{(r, k-r)}(j)}, B^{k-r+[\f{q+1}{2}]}=b^{[\f{q+1}{2}]}], t)\\
&\:\quad=2\int_{\L}\!\dd{a}\,\tilde{\beta}(y_1+y_2,y_1, a, t)
n^{(k+1, \ell-1)}(\x^{k+1}[x^{k+1}= y_1+y_2],\y^{2\ell}_{-(2v-1), -2v}, 
\a^{\ell+1}[a^{\ell+1}=a],\b^{\ell}_{-v}, t) \\
&\:\qquad+ 2\!\sum_{u=1,\neq v}^{k}\!\tilde{\beta}(y_1+y_2, y_1, a^u, t)
n^{(k-1, \ell)}(\x_{-u}^k, \a_{-u}^k, \y^{2\ell}[y^{2v-1}=y_{1}+y_{2}, y^{2v}=x^u], 
\b^{\ell}[b^{v}=a^u], t),
\label{BChighdim2}
\end{aligned}
\end{equation}
%
%%%%%%%%%%%%%%%%%%%%%%%%%%%%%%%%%%%%%%%%%%%%%%%%%%%%%%%%%%%%%%%%%%%%%%%%%%
%
where $\x^{k}_{-u}\!\coloneqq(x^1,...,x^{u-1}, x^{u+1},...,x^k)$,
$\a^{k}_{-u}\!\coloneqq (a^1, ..., a^{u-1}, ...,a^{u+1},... a^k)$, 
$\x^{k}_{-u,
  -v}\!\coloneqq(x^1,...,x^{u-1}\!, x^{u+1}\!,...
,x^{v-1}\!, x^{v+1}\!,\\...,x^k)$,
$\a^{k}_{-u, -v}\!\coloneqq(a^1, ..., a^{u-1}\!, a^{u+1}\!,...,a^{v-1}\!, a^{v+1}\!,...,
a^k)$, $\y^{2\ell}_{-(2v-1), -2v}\!\coloneqq(y^1,...,y^{2v-2}\!\!,
y^{2v+1}\!\!,...,y^{2\ell})$, $\b^{\ell}_{-v}\!\coloneqq(b^1,...,b^{v-1},\\
b^{v+1},...,b^{\ell})$ and $S_k^{-v}\!\coloneqq\{1,2,...,v-1,v+1,..,k\}$.
%$\Omega_r^{-v}$ is the set containing all possible choices of choosing
%r$ elements from $S_k^{-v}$. 
The additional conditions,

\begin{equation}
n^{(k, \ell)}(\x^k,\a^k,\y^{2\ell}, \b^{\ell}, t)=0 \,\, \left\{\begin{array}{l}
\mbox{if any $x_{i}, y_j=0,\infty$} \\
\mbox{if two or more $a_{i}$ or $b_j$=0}
\label{BChighdim0}
\end{array}\right.
\end{equation}
are found by using Eq.~\eqref{quantity} in Eqs.~\eqref{rhobc}.  Note
that if we take $k=1, \ell=0$, with an $m, n$-independent $\beta$, the
``1-point'' total mean population density $n^{(1, 0)}(x, a, t)$ in
volume $x$ and age $a$ at time $t$ is simply
\begin{equation}
n^{(1, 0)}(x, a, t)\!\equiv  
\sum_{m=1}^{\infty}\!\sum_{n=0}^{\infty}\!m\rho_{m, n}^{(1, 0, 0)}(\X^1[X^1=x],\A^1[A^1=a],t)
+\sum_{m=0}^{\infty}\!\sum_{n=1}^{\infty}\!2n \rho_{m, n}^{(0, 1, 0)}(\Y^2_{\rm e}[Y^2=x],\B^1[B^1=a],t),
\end{equation}
and obeys a first-moment (in both dimension and particle number), closed PDE

\begin{equation}
\f{\p n^{(1, 0)}}{\p t} + \f{\p n^{(1, 0)}}{\p a}  +\f{\p (gn^{(1, 0)})}{\p x}
 = -\beta(a, t)n^{(1, 0)}(x, a, t) + \f{1}{2}\f{\p^2 (\sigma^{2}n^{(1, 0)})}{\p x^2}
\label{Meanfield}
\end{equation}
with associated boundary conditions specified at $a=0, x=0, x=\infty$ 
\begin{equation}
\begin{aligned}
n^{(1, 0)}(x, 0, t) & = 2n\sum_{m=0}^{\infty}
\sum_{n=1}^{\infty}\rho_{m, n}^{(0, 1, 0)}(\Y^2_{\rm e}[Y^2=x],\B^1[B^1= 0], t)  \\
\: & =2(m+1)\sum_{m=0}^{\infty}\sum_{n=1}^{\infty}\int_x^{\infty}\!\!\dd{X^1}\!\int_{\L}
\!\dd{A^1}\,\tilde{\beta}(X^1, x, A^1, t)
\rho_{m+1, n-1}^{(1, 0, 0)}(\X^1,\A^1,t)  \\
\: & \hspace{2cm} + 4n\sum_{m=0}^{\infty}\sum_{n=1}^{\infty}\int_x^{\infty}\!\!\dd{Y^2}\!
\int_{\L}\!\dd{B^1}\,\tilde{\beta}(Y^2, x, B^1, t)\rho_{m, n}^{(0, 1, 0)}(\Y^2_{\rm e},B^1, t) \\
\: & = 2\int_x^{\infty}\!\!\dd{z}\int_{\L}\!\dd{a}\,\tilde{\beta}(z, x, a, t)
n^{(1, 0)}(z, a, t), \\
n^{(1, 0)}(0, a, t) & = n^{(1, 0)}(\infty, a, t) = 0.
\label{n1BC}
\end{aligned}
\end{equation}

\begin{figure}[ht]
  \centering
  \includegraphics[width=6.8in]{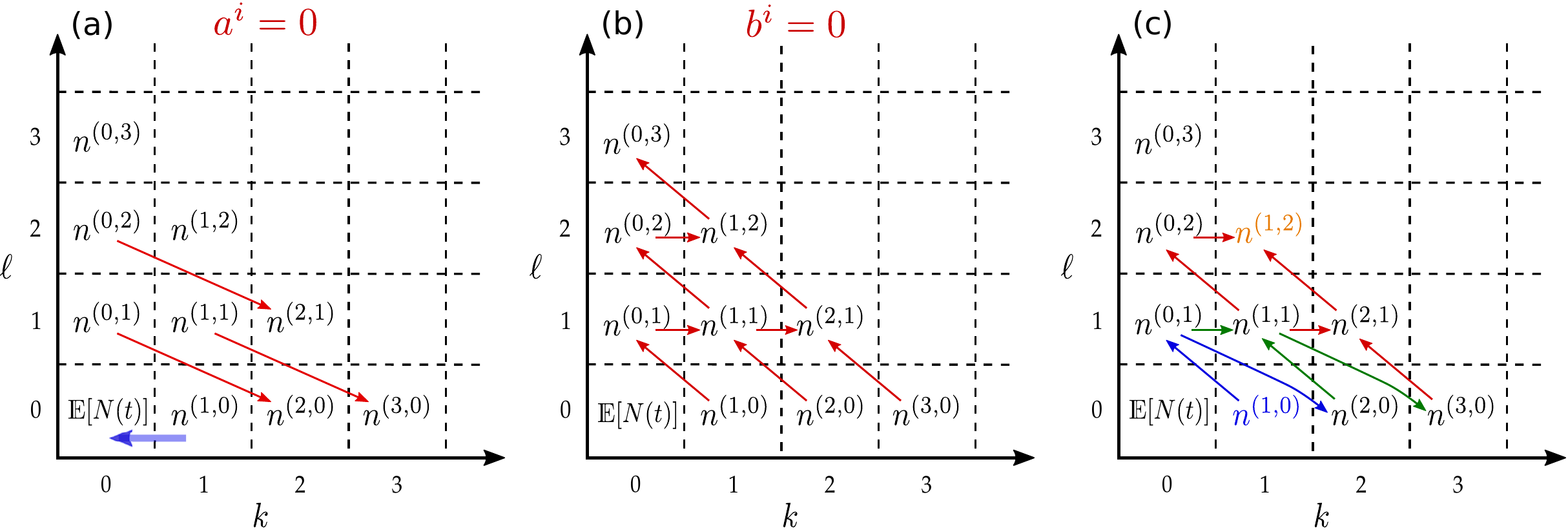}
  \caption{A map of boundary condition interdependences for
    single-density kinetic theory. In (a) we indicate the dependence
    of the boundary condition for $n^{(k, \ell)}(\x^k, \a^k,
    \y^{2\ell}, \b^{\ell}, t)$ if any $a^i=0$. The boundary condition
    for $n^{(k, \ell)}$ depends on itself and $n^{(k-2, \ell+1)}$; for
    example, $n^{(0, 1)}$ is required for the boundary condition for
    $n^{(2, 0)}$, so the red arrow points from $n^{(0,1)}$ to $n^{(2,
      0)}$. In (b) we indicate the dependence of the boundary
    condition for $n^{(k, \ell)}(\x^k, \a^k, \y^{2\ell}, \b^{\ell},
    t)$ if any $b^j=0$. Here, the boundary condition for $n^{(k,
      \ell)}$ depends on $n^{(k+1, \ell-1)}$ and $n^{(k-1, \ell)}$.
    (c) An example of an explicit sequence of calculations to find
    $n^{(1, 2)}$ starting from $n^{(1,0)}$.}
\label{FIG1}
\end{figure}
%
%It is easy to verify that $\prod_{i=1}^k n^{(1)}(x_i, a_i, t)$ solves
%Eqn.~\ref{PDEhighdim} for arbitary $k\in\mathbb{N}^+$.

Note that the PDEs for all multi-point single-density functions
$n^{(k,\ell)}$ are closed. However, the boundary conditions couple
$n^{(k, \ell)}, k+\ell>1$ with $n^{(k+1, \ell-1)}, n^{(k-1, \ell)}$,
or $n^{(k-2, \ell+1)}$.  Thus, although the full models for $n^{(k,
  \ell)}, k+\ell>1$ are not closed, the boundary conditions will only
involve $n^{(k', \ell')}$ such that $k'+2\ell'\leq k+2\ell$, and
therefore all $n^{(k, \ell)}, k+\ell>1$ can be solved sequentially
after we have found $n^{(1, 0)}$. For instance, we can calculate
$n^{(0, 1)}$ from $n^{(1, 0)}$, and then $n^{(2, 0)}$, $n^{(1, 1)}$,
and so on. How the different $n^{(k,\ell)}$ are connected through the
boundary conditions are illustrated in Fig.~\ref{FIG1}, demonstrating
the sequence to follow to fully solve the single-density problem.  The
differential equation satisfied by the lowest order moment
$\mathbb{E}[N(t)]$ requires $n^{(1, 0)}$, as indicated by the shaded
blue arrow in Fig.~\ref{FIG1}(a). In Fig.~\ref{FIG1}(c) we show a
sequence of boundary condition calculations to find $n^{(1, 2)}$: the
equations satisfied by $n^{(1, 0)}$ are fully closed so $n^{(1, 0)}$
can be first calculated. In the second step, we use $n^{(1, 0)}$ to
construct the boundary condition and solve for $n^{(0, 1)}$. The third
step is to use $n^{(0, 1)}$ to construct the boundary condition and
solve for $n^{(2, 0)}$. The boundary condition dependence of $n^{(1,
  0)}, n^{(2, 0)}$ is indicated by blue arrows.  The forth step and
fifth steps are to solve for $n^{(1, 1)}$ and $n^{(3, 0)}$, whose
boundary condition dependences are indicated by the green
arrows. Next, we calculate $n^{(2, 1)}$, $n^{(0, 2)}$, and finally
$n^{(1, 2)}$, whose boundary condition dependences are shown by the
red arrows.

These higher dimensional results capture the stochasticity arising
only from noisy growth of each cell (through the diffusive terms in
Eqs.~\eqref{PDEhighdim} and \eqref{Meanfield}). The demographic
stochasticity arising from random birth (and death) times affects the
total population and is most directly probed by higher number
correlations.  For example, the differential equation satisfied by

\begin{equation}
\mathds{E}[N^2(t)] = \sum_{m=0}^{\infty}\sum_{n=0}^{\infty}(m+2n)^2\rho_{m, n}^{(0, 0, 0)},
\end{equation}
is
\begin{equation}
\begin{aligned}
&\f{\dd \mathds{E}[N^2(t)]}{\dd t} = 
\sum_{m=0}^{\infty}\sum_{n=0}^{\infty}\bigg[(2m^2+4mn+m)\!\int\!\dd{X^1}\dd{A^1}\,
\beta(A^1\!\!,\, t)\rho_{m, n}^{(1, 0, 0)}(\X^1\!\!,\,\A^1\!\!,\, t)\hspace{5mm}\\
&\:\hspace{4.2cm} +(8n^2+4mn+2n)\!\int\!
\dd{Y^2}\dd{B^1}\,\beta(B^1\!\!,\, t)
\rho_{m, n}^{(0, 1, 0)}(\Y^2_{\rm e},B^1\!\!,\, t)\bigg].
\label{N2equation}
\end{aligned}
\end{equation}

The lowest order Eq.~\eqref{Nequation} decouples for $\beta(t)$ which
does not depend on age, trivially reducing to $\dd
\mathds{E}[N(t)]/\dd t = \beta(t) \mathds{E}[N(t)]$. As for
$\mathds{E}[N^2(t)]$, if the division rate function is not dependent
on age, Eq.~\eqref{N2equation} reduces to

\begin{equation}
\f{\dd\mathds{E}[N^2(t)]}{\dd t} = 2\beta(t) \mathds{E}[N^2(t)] +
\beta(t)\mathds{E}[N(t)].
\end{equation}
It is also possible to derive the differential equations satisfied by
any $\dd\mathds{E}[N^k(t)]/\dd t, k\in\mathbb{N}^+$ starting from
Eq.~\eqref{rho}. Such equations, as well as those for higher
number-moments such as $\sum_{m,n} m^k \rho_{m,n}^{(h,k,\ell)}$ are
\textit{not} closed and form complex hierarchies that need additional
assumptions to close.

%%%%%%%%%%%%%%%%%%%%%%%%%%%%%%%%%%%%%%%%%%%%%%%%%%
%%%%%%%%%%%%%%%%%%%%%%%%%%%%%%%%%%%%%%%%%%%%%%%%%%
\section{Generalizations}

\subsection{Incorporation of death}
Here, we show how our kinetic theory is modified 
when an age and size-dependent death, occuring with rate 
$\mu(a, t)$, is incorporated. By defining

\begin{equation}
\gamma(a, t)=\beta(a, t)+\mu(a, t)
\end{equation}
%
%Given a system of $m$ singlets with size $\X^m_{t'}$ and age
%$\A_{t'}^m$ at $t'$ and $n$ doublets of size $\Y^{2n}_{t'}$ and age
%$\B^n_{t'}$, 
%
the joint survival probabilities
 $S_{1, m}$ and $S_{2, n}$ in Eq.~\eqref{fdefinition} are modified by

\begin{equation}
\tilde{S}_{1, m}(t; t', \A^m_{t'}) = 
\prod_{i=1}^m e^{-\int_{t'}^{t}\gamma(A^i_{t'}-t'+s,s)\textrm{ds}},\quad 
\tilde{S}_{2, n}(t; t', \B^n_{t'}) =  
\prod_{j=1}^n\left[e^{-\int_{t'}^{t}\gamma(B_{t'}^j-t'+s,s)\textrm{ds}}\right]^2.
\label{ModJointSurv}
\end{equation}
Following the previous derivations, 
we find

%%%%%%%%%%%%%%%%%%%%%%%%%%%%%%%%%%%%%%%%%%%%%%%%%%%%%%%%%%%%%%%%%%%
\begin{comment}

\begin{equation}
\begin{aligned}
\f{\p \tilde{h}_{m, n}}{\p t} &= 
%-\sum^m_{i=1}\f{\p f_{m, n}}{\p A^i_t} - \sum_{j=1}^{n}\f{\p f_{m, n}}{\p B^j} 
- (\sum_{i=1}^m\gamma(A_t^i, t)+2\sum_{j=1}^n\gamma(B_t^j, t))\tilde{h}_{m, n} - \sum_{i=1}^m\f{\p (\mu(X^i, A^i_t, t)\tilde{h}_{m, n})}{\p X^i_t} -\sum_{j=1}^{2n}
\f{\p (g(Y^j, B^j_t, t)\tilde{h}_{m, n})}{\p Y^j_t} \nonumber \\
&\:\quad+\f{1}{2}\sum_{i=1}^{m}\f{\p^2 (\sigma^2(X^i, A^i_t, t)
\tilde{h}_{m, n})}{(\p X^i_t)^2} + \f{1}{2}\sum_{j=1}^{2n}\f{\p (\sigma^2(Y^j, B^j_t, t)\tilde{h}_{m, n})}{(\p Y^j_t)^2}.
\end{aligned}
\label{Gdeath}
\end{equation}
%
where $\A^m_t = \A^m_{t'} + t-t', \B^n_t=\B^n_{t'}+t-t'$. Likewise, if
we define $\rho_{m, n}(\X^m, \Y^{2n}, \A^m, \B^n, t)$ to be the
probability density function that a system have $m$ singlets of
volumes $\X^m$ with age $\A^m$ and $n$ doublets of volumes $\Y^{2n}$
with age $\B^n$ at time $t$ as defined in Eq.~\eqref{rhodefinition}, by
similar calculation we can check that the differential equation
satisfied by $\rho_{m, n}$ is
\end{comment}
%%%%%%%%%%%%%%%%%%%%%%%%%%%%%%%%%%%%%%%%%%%%%%%%%%%%%%%%%%%%%%%%%%%
%

\begin{align}
\f{\p \rho_{m, n}}{\p t}&+{}\sum^m_{i=1}\f{\p \rho_{m, n}}{\p A^i} + 
\sum_{j=1}^{n}\f{\p \rho_{m, n}}{\p B^j} + 
\sum_{i=1}^m\f{\p (g(X^i, A^i, t)\rho_{m, n})}{\p X^i} +
\sum_{j=1}^{2n}\f{\p (g(Y^j, B^j, t)\rho_{m, n})}{\p Y^j} = \nonumber \\
\: &-\bigg(\sum_{i=1}^m\gamma(A^i, t)+2\sum_{j=1}^n\gamma(B^j, t)\bigg)\rho_{m, n} 
+ \f{1}{2}\sum_{i=1}^{m}\f{\p^2 (\sigma^2(X^i, A^i, t)\rho_{m, n})}{(\p X^i)^2}
+ \f{1}{2}\sum_{j=1}^{2n}\f{\p (\sigma^2(Y^j, B^j, t)\rho_{m, n})}{(\p Y^j)^2} \nonumber \\
\: & + (m+1)\int_{\L^2}\!\!\dd{A^{m+1}}\dd{X^{m+1}}\,\mu(A^{m+1}, t)
\rho_{m+1,n}(\X^{m+1}\!\!\!,\, \Y^{2n}\!\!\!,\, \A^{m+1}\!\!\!,\, \B^{n}\!\!,\, t) \label{RHO_DEATH}\\
\: & +\f{2(n+1)}{m}\sum_{i=1}^m\int_{\L}\!\dd{x}\,\mu(A^i, t)
\rho_{m-1,
  n+1}(\X^m_{-i},\Y^{2n+2}[Y^{2n+1\!\!}=x, Y^{2n+2}=X^i], \A^m_{-i},\B^{n+1}[\B^{n+1}\!\!=A^i], t) \nonumber,
\end{align}
where the argument of $\rho_{m,n}$ in the first two lines is 
$(\X^{m}, \Y^{2n}, \A^{m}, \B^{n}, t)$.

The boundary conditions for $\rho_{m, n}$ are the same as
Eq.~\eqref{BCrho00} and Eq.~\eqref{BCrho} since only cell division
contributes to the boundary term, and no cell can have 0 or infinitely
large volume at any time. Similarly, we can define the marginal
distribution $\rho_{m, n}^{(h, k, l)}(\X^h,\Y_{\rm e}^{2k+2l},\A^h,
\B^k, t)$ and the population density function with respect to volume
$x$ and age $a$ at time $t$ is

\begin{equation}
n^{(1,0)}(x, a, t) = \sum_{m=1}^{\infty}\sum_{n=0}^{\infty}m\rho_{m, n}^{(1,
  0,0)}(\X^1[X^1=x],\A^1[A^1=a],t)+\sum_{m=0}^{\infty}\sum_{n=1}^{\infty}2n \rho_{m,
  n}^{(0, 1,0)}(\Y_{\rm e}^2[Y^2=x],\B^1[B^1=a],t).
\end{equation}
By similar calculations as in Section \eqref{meanfield}, we obtain the
differential equation satisfied by $n^{(1, 0)}(x,a,t)$

\begin{equation}
\f{\p n^{(1,0)}}{\p t} + \f{\p (gn^{(1,0)})}{\p x} 
+ \f{\p n^{(1, 0)}}{\p a} -\f{1}{2}\f{\p^2 (\sigma^2n^{(1,0)})}{(\p x)^2}= -(\beta(a, t)+
\mu(a, t))n^{(1,0)}(x, a, t),
\label{Meanfield_death}
\end{equation}
with boundary conditions specified at $a=0$ and $x=0, \infty$
\begin{equation}
\begin{aligned}
n^{(1,0)}(x,0,t) &= 2n\sum_{m=0}^{\infty}\sum_{n=1}^{\infty}
\rho_{m, n}^{(0, 1,0)}(\Y^2_{\rm e}[Y^2=x],\B^1[B^1= 0], t) \\
\: &=2(m+1)\sum_{m=0}^{\infty}\sum_{n=1}^{\infty}\!\int_x^{\infty}\!\!\dd{X^1}
\int_{\L}\!\dd{A^1}\,\tilde{\beta}(X^1, x, A^1, t)
\rho_{m+1, n-1}^{(1, 0,0)}(\X^1,\A^1,t) \\
\: & \qquad\quad +4n\sum_{m=0}^{\infty}\sum_{n=1}^{\infty}\int_x^{\infty}\!\!\dd{Y^2}
\int_{\L}\!\dd{B^1}\,\tilde{\beta}(Y^2, x, B^1, t)
\rho_{m-1, n}^{(0, 1,0)}(\Y^2_{\rm e},\B^1, t) \\
\: & = 2\int_0^{\infty}\!\!\dd{a}\int_x^{\infty}\! \dd{z}\, 
\tilde{\beta}(z, x, a, t)n^{(1,0)}(z,a,t), \\
n^{(1,0)}(0, a, t) & = n^{(1,0)}(\infty, a, t) = 0.
\label{n1BC_death}
\end{aligned}
\end{equation}

\subsection{Correlated noise in growth rate}
In this subsection we consider a model in which the noise in growth
rates are correlated across cells. By defining $\Z^{m,2n}=(\X^m,
\Y^{2n})$ and $\C^{m, 2n}=(\A^m, B^1, B^1,...,B^n, B^n)$ to be the
volumes and ages of $m$ singlets and $n$ doublets at time $t$, we can
describe the growth rate as
\begin{equation}
\dd{\Z^{m,2n}_t} = G^{m, 2n}(\Z^{m,2n}_t, \C^{m,2n}_t, t)\dd{t} 
+ \Sigma^{m, 2n}(\Z_t^{m,2n}, \C^{m,2n}_t, t)\dd{\W^p_t},
\end{equation}
where $G^{m, 2n}\in\mathbb{R}^{m+2n}$, $\Sigma^{m, 2n}(\Z_t^{m,2n},
\C^{m,2n}_t, t)=(\sigma)_{ij}\in\mathbb{R}^{(m+2n)\times p}$ and
$\W^p_t$ is a $p$-dimensional i.i.d standard Wiener process
\cite{Durrett2005Probability}.  For simplicity, we assume that the
$i^{\textrm{th}}$ component of $G^{m, 2n}$ is $g_i(Z^i_t, C^i_t, t)=g(Z^i, C^i,
t)$, indicating that the deterministic part of the growth rate is
identical for all cells. We further assume that the variance in growth
rates for all cells is identical: $\sum_{\ell=1}^p\sigma_{i,
  \ell}^2=\sigma^2,\, \forall i$.
%
%We define $\rho_{m, n}(\X^m, \Y^{2n}, \A^m, \B^n, t)$ to be the
%probability density function of having $m$ singlets of volumes $\X^m$
%and ages $\A^m$ and $n$ doublets of volumes $\Y^{2n}$ and ages $\B^n$
%at time $t$.
%
Following our derivation in Section \eqref{DERIVATION}, we find
that $\rho_{m, n}(\X^m, \Y^{2n}, \A^m, \B^n,t)$ satisfies

\begin{equation}
\begin{aligned}
\f{\p \rho_{m, n}}{\p t}{}+&\sum_{i=1}^m\f{\p \rho_{m, n}}{\p A^i}+
\sum_{j=1}^n\f{\p \rho_{m, n}}{\p B^i}+ \sum_{i=1}^m
\f{\p(g(t, X^i\!, A^i)\rho_{m, n})}{\p X^i}+\sum_{j=1}^{2n}
\f{\p(g(t, Y^j\!, B^{[(j+1)/2]})\rho_{m, n})}{\p Y^j} = \\
\: &\qquad  -\bigg(\sum_{i=1}^m\beta(A^i\!, t)+ \sum_{j=1}^{n}2\beta(B^j\!, t)\bigg)
\rho_{m, n}(\X^m\!\!,\, \Y^{2n}\!\!\!,\, \A^{m}\!\!,\, \B^{n}\!\!,\,t) +
\sum_{s_1, s_2=1}^{m+2n}\f{1}{2}\f{\p^2 (\rho_{m, n}D_{s_1, s_2})}{\p Z^{s_1}\p Z^{s_2}},
\label{corkinetic}
\end{aligned}
\end{equation}
where $D_{s_1, s_2} = \sum_{\ell=1}^p\sigma_{s_1, \ell}\sigma_{s_2,
  \ell}$. The boundary conditions for $\rho_{m, n}$ are the same as that
described by Eq.~\eqref{BCrho00} and Eq.~\eqref{BCrho}.
%
%Similarly, if we define $\rho_{m, n}$ to be the same as
%Eqn.~\ref{rhodefinition}, the differential equation and boundary
%conditions satisfied by $\rho_{m, n}$ is the same as $p_{m, n}$ %in
%Eqn.~\ref{corkinetic} for $\B^n>0$.
%
Similarly, we can define the marginal distribution density function
$\rho_{m, n}^{(h, k, \ell)}$ in the same way as in Section 3, and it can
be verified that the differential equations as well as the boundary
conditions satisfied by $\rho_{m, n}^{(1, 0, 0)}(\X^1[X^1=x],
\A^1[A^1=a], t), \rho_{m, n}^{(0, 1, 0)}(\X^1[X^1=x],
\A^1[A^1=a], t)$ are the same as those satisfied by $\rho_{m,
  n}^{(1, 0, 0)}(\X^1[X^1=x], \A^1[A^1=a],t)$ and $\rho_{m,
  n}^{(0, 1, 0)}(\Y^1[Y^1=x], \B^1[B^1=a], t)$ in Eq.~\eqref{rho} and
Eq.~\eqref{rhobc}, although the differential equations satisfied by
$\rho_{m,n}$ in Eq.~\eqref{corkinetic} and in Eq.~\eqref{rhodefinition}
are different.  The equation and boundary conditions for the
``1-point'' density function $n^{(1,0)}(x,a,t)$ are identical to those
in Eq.~\eqref{Meanfield} and Eqs.~\eqref{n1BC} since correlations are not
captured by a mean-field description of only one coordinate
($x,a$). The differences between correlated and uncorrelated growth
noise among cells may arise in the differential equations for $n^{(k,
  \ell)}(\x^{k}, \a^{k}, \y^{\ell}, \b^{\ell}, t), \ell+k \geq 2$.

\section{Summary and Conclusions}

In this paper, we rigorously constructed a kinetic theory for
structured populations, in particular for age- and size- structured
cell proliferation models. We considered stochasticity in both an
individual cell's growth rate (``intrinsic'' stochasticity) and the
cell number fluctuations from random birth and death event times
(``demographic'' stochasticity).  Derivations of the kinetic theory
requires separation of 'singlet' and 'doublet' populations, as was
proposed in \cite{chou2016hierarchical}. However, taking into account
both the size and age dependence as well as randomness in growth rates
leads to the much more complex computation which we performed here.

One of our main results are the kinetic equations and boundary
conditions described by Eqs.~\eqref{RHO}, \eqref{BCrho00}, and
\eqref{BCrho}. Marginalized densities are also found to obey more
complex equations that form a hierarchy (Eqs.~\eqref{rho},
\eqref{rhobc}, and \eqref{rhobc0}).  By taking single-density averages
over these equations, we find closed PDEs that govern multi-point
density functions (Eq.~\eqref{PDEhighdim}). However, the associated
boundary conditions, Eq.~\eqref{BChighdim}, couple density functions
of different dimensions.  Nonetheless, density function of all
dimensions can be successively solved starting from the ``1-point''
density $n^{(1,0)}(x,a,t)$ which obeys Eqs.~\eqref{Meanfield} and
\eqref{n1BC}, a 2+1-dimensional second order PDE and boundary
condition that is analogous to the classic McKendrick equation but
that a includes a diffusive size term arising from stochasticity in
growth rates.  The explicit equations for the first and second moments
of the total population are given by Eqs.~\eqref{Nequation} and
\eqref{N2equation}, respectively.

Generalizations and extensions to our basic kinetic theory are also
investigated. For example, we derived the kinetic equations when a
Markovian age-dependent death process is included
(Eqs.~\eqref{RHO_DEATH}, and ~\eqref{Meanfield_death},
\eqref{n1BC_death}).  We also considered noise in growth rates that
are correlated across cells and showed these effects arising in
``cross-diffusion'' terms in the associated kinetic (and higher
moment) equations.

Our unifying kinetic theory enables one to systematically analyze cell
populations at both the individual and population levels.  A full
kinetic theory may be useful for studying other processes such as
failure in multicomponent systems that age and evolve
\cite{SXSUN2018}. Further extensions of our kinetic equations that are
feasible are to include spatial distribution
\cite{auger2008structured} or correlations in growth rates across
\textit{generations} \cite{Xia2020}.  It is also possible to consider
stochasticity for different cell division strategies
\cite{nieto2020continuous}. Finally, efficient numerical methods for
solving our kinetic equations can be developed, for instance in
\cite{Xia2020b} equations similar to Eq.~\eqref{Meanfield} and
Eq.~\eqref{n1BC} which describes the dynamics of $n^{(1, 0)}$ are
solved accurately and efficiently.

\section*{Acknowledgements}
This research was made possible through funding support from the Army
Research Office (W911NF-18-1-0345), the NIH (R01HL146552), and the
National Science Foundation (DMS-1814364).

\section*{References}
\bibliographystyle{iopart-num}
\bibliography{bibliography}

\section*{Appendix: conservation of probability}

%\subsection{Conservation of probability}

We now define probability fluxes 

%\scriptsize
\begin{align}
&J_{m, n;m+1, n-1}(t)= 
(m+1)\int
%_{\L^{m+(2n-2)+m+(n-1)}}
\dd{\X^m}\dd{\Y^{2n-2}}\dd{\A^m}\dd{\B^{n-1}}
\int_{\L^3}\!\dd{y_1}\dd{y_2}\dd{s}\,
\tilde{\beta}_{m+1, n-1}(y_1+y_2, y_1, s, t)\times \nonumber \\[-1pt]
\:& \hspace{5cm} \rho_{m+1, n-1}( \X^{m+1}[\X^{m+1} = y_1+y_2], 
\Y^{2n-2},\A^{m+1}[\A^{m+1}=s],\B^{n-1}, t), \nonumber\\[12pt]
\: & J_{m, n; m-1, n}(t)= \f{2n}{m}\int
%_{\L^{m+(2n-2)+m+(n-1)}}
\dd{\X^m}\dd{\Y^{2n-2}}\dd{\A^m}\dd{\B^{n-1}}
\int_{\L^2}\!\dd{y_1}\dd{y_2}\sum_{i=1}^m\, 
\tilde{\beta}_{m-1, n}(y_1+y_2, y_1, A^i, t)\times \nonumber \\[-1pt]
\:&\hspace{5cm} \rho_{m-1, n}(t, \X^m_{-i},\Y^{2n}[Y^{2n-1}=X^i,Y^{2n}= y_1+y_2], 
\A^m_{-i}, \B^n[B^n=A^i], t), \nonumber \\[12pt]
\:& J_{m, n; m', n'}(t)=0, \quad \textrm{if}\; m+2n-m'-2n'\neq 1.
\label{ProbFlux}
\end{align}
$J_{m, n; m', n'}(t)\dd{t}$ is the probability flux within time $[t,
  t+\dd{t}]$ from state $(m',n')$ to state $(m,n)$ arising from from
cell division. When $\dd{t}$ is sufficiently small, the probability
that more than one cell divides during $[t, t+\dd{t}]$ is $o(\dd{t})$,
which is negligible, allowing us to set $J_{m, n; m', n'}(t)=0$ if
$m+2n-m'-2n'\neq 1$.  We now verify the conservation of probability
flux

\begin{align}
&J_{m-1, n+1; m, n}(t)+J_{m+1, n; m, n}(t) \nonumber \\
&\:\qquad=\int
%_{\L^{m+2n+m+n}}
\dd{\X^m}\dd{\Y^{2n}}\dd{\A^m}\dd{\B^{n}}\bigg(
\sum_{i=1}^m \beta_{m, n}(A^i, t)\rho_{m, n}
%(\X^m, \Y^{2n}, \A^m, \B^{n}, t)
+ \sum_{i=j}^n2 \beta_{m, n}(B^j, t)\rho_{m, n}\nonumber \\
%(\X^m, \Y^{2n}, \A^m, \B^{n}, t))
&\:\qquad= \int
%_{\L^{m+2n+m+n}}
\dd{\X^m}\dd{\Y^{2n}}\dd{\A^m}\dd{\B^{n}} \bigg(
m\beta_{m, n}(A^m, t)\rho_{m, n}
%(\X^m, \Y^{2n}, \A^m, \B^{n}, t)
+2n\beta_{m, n}(B^n, t)\rho_{m, n}\bigg),
%(\X^m, \Y^{2n}, \A^m, \B^{n}, t)).
\label{ProbConse}
\end{align}
where $\rho_{m, n}=\rho_{m, n}(\X^m, \Y^{2n}, \A^m, \B^{n}, t)$.
The first term is

\begin{align}
&J_{m-1, n+1; m, n}(t)=m\int
%_{\L^{(m-1)+2n+(m-1)+n}}
\dd{\X^{m-1}}\dd{\Y}^{2m}\dd{\A^{m-1}}\dd{\B^n}\int_{\L^3}\dd{y_1}\dd{y_2}\dd{A^m}\,
\tilde{\beta}_{m, n}(y_1+y_2, y_1, A^m, t)\times \nonumber\\
\:&\hspace{8cm} \rho_{m, n}(\X^{m}[X^m=y_1+y_2],\Y^{2n},\A^{m},\B^{n}, t) \nonumber\\[9pt]
\:&\quad =m\int
%_{\L^{(m-1)+2n+(m-1)+n}}
\dd{\X^{m-1}}\dd{\Y^{2n}}\dd{\A^{m-1}}\dd{\B^{n}}\int_{\L^2}\dd{A^m}\dd{(y_1+y_2)}
\int_{0}^{y_1+y_2}\dd{y_2}
\tilde{\beta}_{m, n}(y_1+y_2, y_1, A^m, t)\times \nonumber\\
\:&\hspace{8cm} \rho_{m, n}(\X^{m}[X^m=y_1+y_2],\Y^{2n},\A^{m},\B^{n}, t) \nonumber\\[9pt]
\:& \quad=m\int
%_{\L^{(m-1)+2n+(m-1)+n}}
\dd{\X^{m-1}}\dd{\Y^{2n}}\dd{\A^{m-1}}\dd{\B^{n}}\int_{\L^2}\dd{A^m}\dd{X^m}\,
\beta_{m, n}(A^m, t)\rho_{m, n}
%(\X^{m}, \Y^{2n},\A^{m},\B^{n},t))\\
%\:& \quad=m\int_{\L^{m + 2n + m + n}}
%\dd{\X^m}\dd{\Y^{2n}}\dd{\A^m}\dd{\B^n}
%(\beta_{m, n}(A^{m}, t)\rho_{m, n}(\X^{m}, \Y^{2n},\A^{m},\B^{n}, t)).
\end{align}
which is exactly the first term on the right hand side of
Eq.~\eqref{ProbConse}. The second term

\begin{align}
&J_{m+1, n; m, n}(t)=\f{2n}{m+1}\int
%_{{\L}^{(m+1)+(2n-2)+(m+1)+(n-1)}}
\dd{\X^{m+1}}\dd{\Y^{2n}}\dd{\A^{m+1}}\dd{\B^{n-1}}\int_{\L^2}\dd{y_1}\dd{y_2}
\sum_{i=1}^{m+1}\tilde{\beta}_{m, n}(y_1+y_2, y_1, A^i, t)\times\nonumber\\
&\hspace{5cm}\rho_{m, n}(\X^{m+1}_{-i}, \Y^{2n}[Y^{2n-1}=X^i,Y^{2n}=y_1+y_2],
\A^{m+1}_{-i},\B^n[B^n=A^i], t)\nonumber\\[9pt]
&\:\quad=\f{2n}{m+1}\sum_{i=1}^{m+1}\int
%_{\L^{(m+1)+(2n-2)+(m+1)+(n-1)}}
\dd{\X^{m+1}}\dd{\Y^{2n-2}}\dd{\A^{m+1}}\dd{\B^{n-1}}
\int_{\L}\dd{(y_1+y_2)}\int_{0}^{y_1+y_2}\dd{y_1}\,
\tilde{\beta}_{m, n}(y_1+y_2, y_1, A^i, t)\times\nonumber\\
&\hspace{5cm} \rho_{m, n}(\X^{m+1}_{-i}, \Y^{2n}[Y^{2n-1}=X^i, Y^{2n}=y_1+y_2],
\A^{m+1}_{-i}, \B^n[B^n=A^i],t)\nonumber\\[9pt]
&\:\quad=2n\int
%_{\L^{m+2n+m+n}}
\dd{\X}^{m}\dd{\Y}^{2n}\dd{\A}^{m}\dd{\B}^{n}\, \beta_{m, n}(B^n, t)\rho_{m, n}
%(\X^{m},\Y^{2n}, \A^{m},\B^n, t))\\
\end{align}
which is precisely the second term on the right hand side of
Eq.~\eqref{ProbConse}. We have thus verified that the probability flux
out of state ($m,n$) due to cell division is the sum of probability
currents into ($m-1, n+1$) and into ($m+1, n$).  Summing up over $m$
and $n$, we obtain for $m+n>0$

\begin{align}
&\sum_{m=0}^{\infty}\sum_{n=0}^{\infty}\bigg(J_{m-1, n+1; m, n}(t)
+J_{m+1, n; m, n}(t)\bigg) = \nonumber \\
& \hspace{3cm} \sum_{m=0}^{\infty}\sum_{n=0}^{\infty}\int
%_{{\L}^{m+2n+m+n}}
\dd{\X^m}\dd{\Y^{2n}}\dd{\A^m}\dd{\B^n}\, \bigg(m\beta_{m, n}(A^m, t)\rho_{m, n}
%(\X^{m},\Y^{2n}, \A^{m},\B^n, t)\\
+2n\beta_{m, n}(B^n, t)\rho_{m, n}
%(\X^{m},\Y^{2n}, \A^{m},\B^n, t)
\bigg).
\end{align}
Finally, it is readily observed that 

\begin{align}
&\sum_{m=0}^{\infty}\sum_{n=0}^{\infty}\int
%_{\L^{m+2n+m+n}}
\dd{\X^m}\dd{\Y^{2n}}\dd{\A^m}\dd{\B^n}\f{\p \rho_{m, n}}
%(\X^m, \Y^{2n}, \A^m, \B^n, t)
{\p t}=   \nonumber \\
&\qquad \quad \sum_{m=0}^{\infty}\sum_{n=0}^{\infty}\sum_{j=1}^n\int
%_{{\L}^{m+2n+m+n-1}}
\dd{\X^m}\dd{\Y^{2n}}\dd{\A^m}\dd{\B^{n}_{-j}}
\rho_{m, n}(\X^m, \Y^{2n}, \A^m, \B^n[B^j=0], t)   \nonumber \\
&\:\qquad\qquad\qquad\quad  -\sum_{m=0}^{\infty}\sum_{n=0}^{\infty}\int
%_{{\L}^{m+2n+m+n}}
\dd{\X^m}\dd{\Y^{2n}}\dd{\A^m}\dd{\B^n}
\bigg(\sum_{i=1}^m\beta_{m, n}(A^i, t)\rho_{m, n}
%
%(\X^{m},\Y^{2n}, \A^{m},\B^n, t)
%
+\sum_{j=1}^n2\beta_{m, n}(B^j, t)\rho_{m, n}\bigg) \nonumber \\
%
%(\X^{m},\Y^{2n}, \A^{m},\B^n, t))\\
&\:\qquad \qquad  =\sum_{m=1}^{\infty}\sum_{n=0}^{\infty}(J_{m, n;m-1, n}-J_{m-1, n+1; m, n})
-\sum_{m=0}^{\infty}\sum_{n=0}^{\infty}J_{m+1, n;m, n}+
\sum_{m=0}^{\infty}\sum_{n=1}^{\infty}J_{m, n; m+1, n-1} = 0
%
%-\sum_{m=1}^{\infty}\sum_{n=0}^{\infty}J_{m-1, n+1; m, n}=0.
\end{align}
Therefore, we have verified that

\begin{align*}
\sum_{m=0}^{\infty}\sum_{n=0}^{\infty}\int
%_{\L^{m+2n+m+n}}
\dd{\X^m}\dd{\Y^{2n}}\dd{\A^m}\dd{\B^n}\rho_{m, n}
%(\X^m, \Y^{2n}, \A^m, \B^n, t) 
%= \sum_{m=0}^{\infty}\sum_{n=0}^{\infty}\int
%_{\L^{m+2n+m+n}}
%\dd{\X^m}\dd{\Y^{2n}}\dd{\A^m}\dd{\B^n}\rho_{m, n}(\X^m, \Y^{2n}, \A^m, \B^n, 0).
\end{align*}
is time-independent.

%%%%%%%%%%%%%%%%%%%%%%%%%%%%%%%%%%%%%%%%%%%%%%%%%%%%%%%%%%%%%%%%%%%%%%%%%
\end{document}